\documentclass[12pt]{article}
\usepackage[latin1]{inputenc}
\usepackage[a4paper,left=2.5cm,right=2.5cm,top=2cm,bottom=2cm,footnotesep=.75cm,headheight=13.6pt]{geometry}
\usepackage{setspace}
\usepackage{amsthm}
\usepackage{amsmath}
\usepackage{amsfonts}
\usepackage{amssymb}
\usepackage{graphicx}
\usepackage[small]{caption}
\usepackage{authblk}
\usepackage{rotating}
\usepackage{float}
\usepackage{pdflscape}
\usepackage{subcaption}
\usepackage{longtable}
\usepackage{xcolor}
\usepackage{ulem}

\title{\Large{Mean Square Errors of factors extracted using principal components, linear projections, and Kalman filter}}

\author[1]{\small Matteo Barigozzi}
\author[2]{\small Diego Fresoli}
\author[3]{\small Esther Ruiz\thanks{Corresponding author. E-mail: ortega@est-econ.uc3m.es}}
\affil[1]{\footnotesize Department of Economics, University of Bologna (Italy)}
\affil[2]{Department of Economic Analysis-Quantitative Economics, Universidad Autonoma de Madrid (Spain)}
\affil[3]{Department of Statistics, Universidad Carlos III de Madrid (Spain)}

\date{\small \today}

\begin{document}

\maketitle

\begin{abstract}
Factor extraction from systems of variables with a large cross-sectional dimension, $N$, is often based on either Principal Components (PC)-based procedures, or Kalman filter (KF)-based procedures. Measuring the uncertainty of the extracted factors is important when, for example, they have a direct interpretation and/or they are used to summarized the information in a large number of potential predictors. In this paper, we compare the finite $N$ mean square errors (MSEs) of PC and KF factors extracted under different structures of the idiosyncratic cross-correlations. We show that the MSEs of PC-based factors, implicitly based on treating the true underlying factors as deterministic, are larger than the corresponding MSEs of KF factors, obtained by treating the true factors as either serially independent or autocorrelated random variables. We also study and compare the MSEs of PC and KF factors estimated when the idiosyncratic components are wrongly considered as if they were cross-sectionally homoscedastic and/or uncorrelated. The relevance of the results for the construction of confidence intervals for the factors are illustrated with simulated data. 
\end{abstract}



\setcounter{page}{1}

\newpage

\doublespacing

\section{Introduction}

Currently, large systems of macroeconomic and/or financial time series are easily accessible.  In this context, the main goal of Dynamic Factor Models (DFMs), originally introduced by Geweke (1977) and Sargent and Sims (1977),  is to explain the dynamics of these systems using a reduced number of underlying unobservable common factors. The DFMs we focus on are based on the premise that the contemporaneous covariation among the time series within a particular system of interest can be traced to these latent factors.

Among the procedures for factor extraction available in the literature, Principal Components (PC)-based procedures are very popular in empirical applications. PC-based procedures are non-parametric in the sense that they do not assume any particular distribution of the underlying factors and idiosyncratic components. In fact,  PC-based procedures implicitly treat the factors as if they were fixed, i.e., non-stochastic, variables.

Alternatively, factors can be extracted using Kalman filtering 
(KF)-based procedures, which treat the underlying factors as unobserved stochastic variables that can have some kind of temporal dependence; see the recent survey by Ruiz and Poncela (2022). Finally, one can also assume that the factors are stochastic and serially independent and identically distributed (\textit{iid}). In this case, KF boils down to a static Linear Projection (LP).

In many important empirical implementations of DFMs, the factors are of interest in themselves, being  often relevant for  econometricians and policy decision makers; see, for example, Stock and Watson (2017). This is the case when, for example, the factors are used to construct indexes of economic activity, of macroeconomic uncertainty, or of systemic financial volatility; see  Henzel and Rengel (2017), Greenaway-McGrevy \textit{et al}. (2018) and Lewis \textit{et al.} (2022), for some selected empirical applications with the factors representing indexes. 
Furthermore, the factors can be used to construct scenarios for economic activity as in Gonz\'{a}lez-Rivera, Maldonado and Ruiz (2019) and Gonz\'{a}lez-Rivera, Rodr\'{i}guez-Caballero and Ruiz (2024), who identify the factors with underlying economic conditions and use their uncertainty to construct economic scenarios for economic growth. Also, the uncertainty of the factors is relevant when dealing with diffusion models in which the factors summarize the information contained in a large number of predictors; see, for example, Bai and Ng (2006a) and Lewis \textit{et al}. (2022) for the construction of prediction intervals. Similarly, in the context of factor-augmented quantile regressions, Ando and Tsay (2011) and Amburgey and McCracken (2022) propose an information criteria and an out-of-sample performance test, respectively,  in which the uncertainty of the factors is taken into account. Finally, the uncertainty of the factors is also relevant when testing whether some observable economic variables are in fact the underlying factors as in Bai and Ng (2006b). 

All the above mentioned applications require measuring uncertainty of the estimated factors for inference and/or to establish the statistical legitimacy of the results.
The uncertainty of the factors is typically measured using the finite sample Mean Squared Error (MSE) which, when multiplied by $N$, provides a finite sample estimator of the asymptotic covariance of the estimated factors.

In this paper, we compare the MSEs of the factors extracted by PC-based, LP-based and KF-based procedures, under different structures of the idiosyncratic covariance matrix.  Specifically, we consider both the correctly specified case in which the idiosyncratic covariance matrix is full, and the mis-specified cases in which the  idiosyncratic covariance matrix is treated as if it were diagonal or even spherical.
Since in practice considering the correctly specified estimators requires  the hard, if not unfeasible, task of inverting the full idiosyncratic covariance matrix, the mis-specified cases are the relevant ones.

In order to simplify the comparison of MSEs avoiding the interference of the properties of the parameter estimators, we compare them assuming that the factor loadings, the cross-sectional idiosyncratic variances and covariances, and any possible parameter determining the dynamics of the factors are known, an approximation which is always valid provided that the temporal sample size, $T$, is such that $\sqrt N/T\to 0$ as $N,T\to\infty$, with $N$ being the cross-sectional dimension of the system (see Section \ref{sec:param} for details). In this case PC-based methods become Least Squares (LS)-based methods and we refer to them as such. When seeing the PC-based estimators as LS-based estimators, it becomes clear that they are implicitly treating the factors as deterministic, or equivalently they are estimating a realization of the factors. This is different from the case of LP-based and KF-based factor extraction methods, which are based on treating the factors as random variables.

Several important conclusions relevant for empirical researchers interested in the uncertainty of underlying factors extracted from large systems of economic/financial time series are obtained. First, we show that regardless of the particular estimator considered, the MSE matrix of the factors is only diagonal when the factors are extracted using the true structure of the idiosyncratic variances and covariances. Extracting the factors wrongly assuming that the idiosyncratic covariance matrix is diagonal/spherical generates non-zero correlations between them. 

Second, for each type of factor extraction procedures considered, those based on assuming full idiosyncratic covariance matrices are never less efficient than those obtained assuming a diagonal covariance, which, in turn, are more efficient than those obtained assuming that it is spherical. However, note that extracting the factors assuming that the idiosyncratic covariance matrix is full is a hard problem in practice due to problems associated to the estimation of such large covariance matrix. 

Third, the difference between the MSE of the factors extracted using LP-based procedures and that of the corresponding LS-based factors is negative semidefinite. This difference is due to the fact that the former procedures treat the factors as random and estimate their conditional means, while the latter ones estimate a realization of the factors, consequently having larger uncertainty. Furthermore, the difference between the MSE of the factors extracted using KF-based procedures and that of the factors extracted by the corresponding LP-based procedures is small unless the persistence of the factors is large, which is the relevant case in many empirical applications. 

Fourth, regardless of the particular structure of the idiosyncratic covariance matrix, the MSEs of all nine factor extraction procedures considered in this paper, converge to a matrix of zeros as $O(N^{-1})$. The factors extracted by all procedures are $\sqrt N$-consistent and coincide asymptotically and so their MSEs (multiplied by $N$) coincide asymptotically. 


In summary, according to our results, our recommendation to the empirical researcher concerned with factor extraction efficiency is that if $N$ is small or moderate the factors should be extracted by KF allowing for idiosyncratic heteroscedasticity. The reduction in MSE of KF with respect to using LP is relevant when the factors are highly persistent.

While this is the first paper to investigate the finite sample properties of the MSE of various factor estimators, many papers have studied their asymptotic properties. The asymptotic distribution of the PC estimator of the factors is derived by Bai (2003) and extended to Weighted Principal Components (WPC) by Breitung and Tenhofen (2011), with weights given by the PC estimator of the idiosyncratic variances, thus addressing cross-sectional heteroscedasticity. The same distribution of the WPC is derived by Bai and Li (2016) when the loadings and idiosyncratic variances are estimated via Quasi-Maximum Likelihood (QML). Choi (2012) derives the asymptotic distribution when taking into account the whole structure of the idiosyncratic covariance matrix and proposes a Generalized Principal Components (GPC) estimator, which, however, is infeasible since the sample idiosyncratic covariance estimated via PC is always singular and cannot be inverted. Recently, Mao \textit{et al}. (2024) derive the asymptotic distribution of the factors when they are assumed to be \textit{iid} and extracted using LP with the loadings estimated by QML. The latter is asymptotically equivalent to the WPC; see Bai and Li (2016).  Finally, when the factors are extracted using KF techniques, their asymptotic distribution is derived in Barigozzi and Luciani (2024).

The rest of the paper is organized as follows. Section \ref{sec:DFM} describes the main assumptions considered in this paper. Section \ref{sec:PC} describes the MSEs of the factors extracted using LS-based procedures. Section \ref{sec:KFS} deals with LP- and KF-based factor extraction. Section \ref{sec:param} discusses the case of estimated parameters.
Simulations are carried out in Section \ref{sec:Simu} to illustrate the results. Section \ref{sec:final} concludes.

\section{The dynamic factor model}
\label{sec:DFM}

Denote by $\mathbf{Y}_t$ the stationary zero mean $N \times 1$ vector of observations at time $t=1,\ldots,T$, represented by the following  DFM
\begin{equation}
\label{eq:DFM}
\mathbf{Y}_t=\mathbf{\Lambda} \mathbf{F}_t+ \boldsymbol{\varepsilon}_t,
\end{equation}
where $\mathbf{\Lambda}$ is the $N\times r$ matrix of factor loadings, $\mathbf{F}_t=\left(  F_{1t},\ldots,F_{rt} \right) ^{\prime}$ is the $r \times 1$ vector of underlying unobserved factors, and $\boldsymbol{\varepsilon}_t= \left( \varepsilon_{1t},\ldots, \varepsilon_{Nt} \right)^{\prime}$ is the $N \times 1$ vector of idiosyncratic components with finite positive-definite symmetric covariance matrix $\mathbf{\Sigma}_{\varepsilon}$. Since the matrix $\mathbf{\Sigma}_{\varepsilon}$ is allowed to be full we call model (\ref{eq:DFM}) an approximate factor model as opposed to an exact one where $\mathbf{\Sigma}_{\varepsilon}$ would be diagonal. As it is common in this literature, the number of factors, $r \ll N$, is assumed to be fixed and independent of $N$.

The following assumptions are made about the loadings, factors and idiosyncratic components in model (\ref{eq:DFM}):\footnote{Hereafter, if $\mathbf{A}$ is a real $n\times m$ matrix, $\Vert \mathbf{A} \Vert $ denotes  the spectral norm of $\mathbf{A}$, i.e., $\Vert \mathbf{A} \Vert = \sqrt{ \nu^{(1)}(\mathbf{AA}^{\prime})}$, where $\nu^{(k)}(\mathbf{B})$ is the $k$-th largest eigenvalue of square matrix $\mathbf{B}$. The norm of a generic $n$-dimensional vector $\mathbf{v}=\left(v_1,\ldots,v_n \right)^{\prime}$ is given by $\Vert \mathbf{v} \Vert = \sqrt{\sum_{i=1}^{n} v_i^2}$.}

\begin{enumerate}

\item Loadings

\begin{enumerate}
\item $\lim_{N\rightarrow\infty} \Vert \frac{ \mathbf{\Lambda}^{\prime} \mathbf{\Lambda}}{N}-\mathbf{\Sigma}_{\Lambda}\Vert =0$, where $\mathbf{\Sigma}_{\Lambda}$ is an $r \times r$ finite positive definite matrix.

\item  $\forall i=1,\ldots,N, \Vert \boldsymbol{\lambda}_{i}\Vert\leq M_{\Lambda}$, for some finite positive real $M_{\Lambda}$, where $\boldsymbol{\lambda}_{i}^\prime$ is the $i$th row of $\mathbf{\Lambda}$.

\item $\lim_{N\rightarrow\infty} \Vert \frac{\mathbf{\Lambda}^{\prime} \mathbf{\Sigma}_{\varepsilon}^{-1} \mathbf{\Lambda}}{N}-\mathbf{\Sigma}^{*}_{\Lambda}\Vert =0$, where $\mathbf{\Sigma}^{*}_{\Lambda}$ is an $r \times r$ positive definite matrix, and $\mathbf{\Sigma}_{\varepsilon}=E(\boldsymbol{\varepsilon}_t\boldsymbol{\varepsilon}_t^\prime)$.

\item $\lim_{N\rightarrow\infty} \Vert \frac{\mathbf{\Lambda}^{\prime} \mathbf{\Sigma}_{\varepsilon} \mathbf{\Lambda}}{N}-\mathbf{\Sigma}^{\dagger}_{\Lambda}\Vert =0$, where $\mathbf{\Sigma}^{\dagger}_{\Lambda}$ is an $r \times r$ positive definite matrix.

\item $\lim_{N\rightarrow\infty} \Vert \frac{\mathbf{\Lambda}^{\prime} \mathbf{\Sigma}^{*-1}_{\varepsilon} \mathbf{\Lambda}}{N}-\mathbf{\Sigma}^{**}_{\Lambda}\Vert =0$ , where $\mathbf{\Sigma}^{**}_{\Lambda}$ is an $r \times r$ positive definite matrix, and $\mathbf{\Sigma}^{*}_{\varepsilon}=\mathrm{diag}(\mathbf{\Sigma}_{\varepsilon})$.

\item $\lim_{N\rightarrow\infty} \Vert \frac{\mathbf{\Lambda}^{\prime} \mathbf{\Sigma}^{*-1}_{\varepsilon} \mathbf{\Sigma}_{\varepsilon} \mathbf{\Sigma}^{*-1}_{\varepsilon}  \mathbf{\Lambda}}{N}-\mathbf{\Sigma}^{\ddag}_{\Lambda}\Vert =0$ , where $\mathbf{\Sigma}^{\ddag}_{\Lambda}$ is an $r \times r$ positive definite matrix.
\end{enumerate}

\item Factors. We consider different assumptions depending of whether the factors are assumed to be fixed unknown constants or they are assumed to be stochastic:\footnote{See, for example, Bai and Li (2012, 2016), Onatski (2012), Breitung and Eickmeier (2016), and Freyaldenhoven (2022), for factors treated as fixed unknown constants.}


i)  Deterministic factors: $\{\mathbf F_t,\, t=1,\ldots ,T\}$ is a collection of fixed constants characterized by the following properties.

\begin{enumerate}

\item $\forall t=1,\ldots,T, \Vert \mathbf{F}_t \Vert \leq c$,  for some finite positive real $c$.

\item Let $\mathbf{M}_{FF}= \frac{1}{T} \sum_{t=1}^T \left( \mathbf{F}_t-\bar{\mathbf{F}} \right) \left( \mathbf{F}_t - \bar{\mathbf{F}} \right) ^{\prime}$, where $\bar{\mathbf{F}}=\frac{1}{T} \sum_{t=1}^T \mathbf{F}_t$, and
 $\lim_{T\rightarrow\infty} \Vert\mathbf{M}_{FF}- \bar{\mathbf{M}}_{FF}\Vert=0$ where $\bar{\mathbf{M}}_{FF}$ is an $r \times r$ finite positive definite matrix.

\end{enumerate}

ii)  Stochastic factors: $\{\mathbf F_t, t\in\mathbb Z\}$ is a stochastic process characterized by the following properties.

\begin{enumerate}
\item $\forall t=1,\ldots,T, E(\mathbf{F}_t)=0$ and $E(\mathbf{F}_t \mathbf{F}^{\prime}_t)=\mathbf{\Sigma}_F$, which is an $r\times r$ finite positive definite matrix. 

\item $\forall t=1,\ldots,T, E(\Vert \mathbf{F}_t\Vert^4)< K_F$, for some finite positive real $K_F$.

\item $\mathrm{plim}_{T\rightarrow \infty} \| \frac{1}{T} \mathbf{F}^{\prime} \mathbf{F} - \mathbf{\Sigma}_{F}\|=0$, where $\mathbf{F}=(\mathbf{F}_1,\ldots,\mathbf{F}_T)^{\prime}$ is the $T \times r$ matrix of factors.

\end{enumerate}


\item Idiosyncratic components

\begin{enumerate}
\item $\boldsymbol{\varepsilon}_t$ is white noise. 

\item $\mathbf{\Sigma}_{\varepsilon}$ is such that, for $i\neq j$, $\left| E(\varepsilon_{it} \varepsilon_{jt}) \right| \leq M_{ij}$, where $M_{ij}$ are positive real numbers such that 
$\sum_{\substack{j=1\\i\neq j}}^N M_{ij}\leq M_{\varepsilon}$
, for some positive real $M_{\varepsilon}$, independent of $N$.

\item $\forall i=1,\ldots,N$ and $t=1,\ldots,T, E[\varepsilon_{it}^4]\leq Q_{\varepsilon}$ for some finite positive real $Q_{\varepsilon}$.

\item $\forall N,T$ and $j=1,\ldots,N$, $E\left[ \left(  \frac{1}{\sqrt{NT}} \sum_{i=1}^N \sum_{t=1}^T \left[  \varepsilon_{it} \varepsilon_{jt} - E(\varepsilon_{it} \varepsilon_{jt}) \right] \right)  ^2 \right] < K_{\varepsilon}$, for some positive real $K_{\varepsilon}$.

\item The idiosyncratic components and the factors are independent at all leads and lags.

\end{enumerate}

\end{enumerate}

According to the assumptions above, the matrix of loadings is non-random having asymptotically maximum column rank, $r$, and, for any given cross-sectional dimension, $N$, each factor has a finite contribution to each observed series. Also, note that, although assumptions 1(c) and 1(d) on the loadings can be derived from 1(a) and 3(b), they are useful to define the matrices $\mathbf{\Sigma}_\varepsilon^*$ and $\mathbf{\Sigma}_\varepsilon^{\dagger}$. We also assume that the explanatory power of the common components strongly dominates the explanatory content of the idiosyncratic components, that is the factors are pervasive; see Barigozzi and Hallin (2025) for a detailed description of fundamental issues in the theory and practice of factor models.

Furthermore, if the factors are fixed, their norms are uniformly bounded over time, and their sample covariance matrix converges to a positive definite limit as the temporal dimension grows. In contrast,  if the factors are stochastic, they are not degenerated: they have zero mean, a positive covariance matrix, finite fourth-order moments, and their sample covariance matrix is a $\sqrt{T}$-consistent estimator of its population counterpart. 

Turning to the idiosyncratic components, they have finite fourth order moments and covariance $\mathbf{\Sigma}_{\varepsilon}$ assumed to be positive definite and with bounded eigenvalues not only for every finite cross-sectional dimension $N$, but also in the limit when $N\rightarrow \infty$. Indeed, assumption 3(b) bounds the $L_1$ norm and consequently, also the $L_2$ norm. Therefore, by the Perron-Frobenius theorem, the largest eigenvalue of $\mathbf{\Sigma}_{\varepsilon}$ is also bounded.  Hence, the idiosyncratic components are allowed to be weakly cross-sectionally correlated and heteroscedastic. It follows also that $\mathbf{\Sigma}_{\varepsilon}^{-1}$ exits and is finite even in the limit $N\to\infty$. 

Finally, note also that, even if it not necessary for the theory in this paper, we assume serially uncorrelated idiosyncratic components for simplicity; see Boivin and Ng (2006) and Luciani (2014), who show that the relevant element when forecasting using DFMs is the forecast of the common component, but not the idiosyncratic one. This assumption is also made by Choi (2012) and Bai and Li (2012), among others. 
Moreover, since here we are interested in factors (and not loadings) estimation, the relevant dimension is the cross-sectional one, meaning that all estimators considered are studied at a fixed point in time $t$, and, clearly, in that case serial dependence of the idiosyncratic components is not relevant. 

Throughout, given our interest on the factors only, we consider the loadings in $\mathbf{\Lambda}$ and variances and covariances in $\mathbf{\Sigma}_{\varepsilon}$, and any possible parameter determining the dynamics of the factors, as known. This simplifies our treatment and allows us to focus on the main differences among the considered estimators. As shown in Section \ref{sec:param}, treating the parameters as known is harmless as long as we estimate them consistently (via PC or QML) and the sample size, $T$ is such that $\sqrt N/T\to 0$ as $N,T\to\infty$. 

Now, although, for  known $\mathbf{\Lambda}$ and $\mathbf{\Sigma}_{\varepsilon}$,  the factors can be identified and, thus, estimated without imposing any restriction, 
we still follow the usual approach of the literature and we impose $r^2$ restrictions which would allow us to identify both the factors and the loadings, were the latter unknown.
 There are two main reasons for this strategy. First, under the imposed restrictions, our results can be directly extended and compared with those obtained when the loadings and idiosyncratic variances and covariances are unknown and must be estimated. Second, some of our derivations are considerably simplified by imposing the restrictions given below. Specifically, in this paper, we assume that (see, e.g., Bai and Li, 2012, 2016, and  Bai and Ng, 2013)
\begin{enumerate}
\item [(I)] $\frac{1}{T}\mathbf{F}^{\prime}\mathbf{F}=\mathbf{I}_r$ for all $T$, which amounts to $\frac{r(r+1)}{2}$ restrictions.
\item [(II)] $\frac{1}{N}\mathbf{\Lambda}^{\prime} \mathbf{\Sigma}_{\varepsilon}^{-1}\mathbf{\Lambda}$ is diagonal (with distinct elements in the main diagonal arranged in decreasing order) for all $N$, which amounts to  $\frac{r(r-1)}{2}$ restrictions. 
\end{enumerate}

About (I), note that such restriction does not make sense in the case of stochastic factors. Consequently, all results derived should be considered as holding either for deterministic factors or as holding conditional on a specific realization of the stochastic factors. About (II), this is the typical restriction imposed in QML estimation, while the usual restriction for PC factor extraction is that $\frac 1N\mathbf{\Lambda}^\prime\mathbf{\Lambda}$ is diagonal for all $N$.

Finally, hereafter, for a given estimator $\mathbf f_t$ of the factors $\mathbf F_t$ the MSE of $\mathbf f_t$ is defined as the $r\times r$ matrix:
\[
\text{MSE}(\mathbf f_t)=E[(\mathbf{f}_t-\mathbf F_t)(\mathbf{f}_t-\mathbf F_t)'].
\]

\section{Least Squares}
\label{sec:PC}

In this section, we consider the factors as fixed deterministic constants and we analyse the MSE of the estimated factors extracted using LS-based procedures under different structures of the idiosyncratic covariance matrix. All results can be directly extended to stochastic factors by considering the MSEs and asymptotic distributions derived as conditional on a specific realization of the factors.

If the loadings, $\mathbf{\Lambda}$, and the elements of the idiosyncratic covariance matrix, $\mathbf{\Sigma}_{\varepsilon}$, were known, the DFM in (\ref{eq:DFM}) can be seen as a seemingly unrelated regression equations (SURE) model.  In this case, at each time period, $t$, the most efficient estimator of the factors coincides with the following Generalized Least Squares (GLS) estimator (see Zellner, 1962,  and Bai and Liao, 2016)
\begin{equation}
\label{eq:GLS}
\mathbf{f}^{GLS}_t=\left( \mathbf{\Lambda}^{\prime} \mathbf{\Sigma}^{-1}_{\varepsilon} \mathbf{\Lambda} \right) ^{-1} \mathbf{\Lambda}^{\prime} \mathbf{\Sigma}^{-1}_{\varepsilon} \mathbf{Y}_t.
\end{equation}
Notice that, in the case of fixed factors, the GLS is also the Maximum Likelihood Estimator (MLE) of the factors, which here are just fixed parameters  (see Anderson and Rubin, 1956). This is to be expected since indeed the MLE of a SURE is GLS. 

Note that if we compute the GLS in \eqref{eq:GLS} using the PC estimator of the loadings and a consistent estimator of the idiosyncratic covariance, then we would obtain the GPC estimator studied by Choi (2012). However, since the estimated idiosyncratic covariance matrix obtained from PC is singular by construction, i.e., it is not invertible, we would need a regularized estimator, such as, e.g., the sparse estimators studied by Fan {\it et al.} (2013) or Bai and Liao (2016).

From \eqref{eq:GLS} we can directly derive the finite sample MSE of $\mathbf{f}_t^{GLS}$:
\begin{align}
\label{eq:MSE_1}
\text{MSE}(\mathbf{f}^{GLS}_t)
&=(\mathbf{\Lambda}^{\prime} \mathbf{\Sigma}_{\varepsilon} ^{-1} \mathbf{\Lambda})^{-1}.
\end{align}
Several interesting insights can be obtained about the finite sample MSE in (\ref{eq:MSE_1}). First, note that, by our identifying restrictions, $\mathbf{\Lambda}^{\prime} \mathbf{\Sigma}_{\varepsilon} ^{-1} \mathbf{\Lambda}$ is diagonal and, consequently, the estimated factors are mutually uncorrelated. 

Second, the identifying restrictions also imply that $\mathbf{\Lambda}^{\prime} \mathbf{\Sigma}_{\varepsilon} ^{-1} \mathbf{\Lambda}$ represents the ``signal-to-noise-ratio'' of the common components and, therefore, their strength when compared with the idiosyncratic components. As expected, the uncertainty associated with the factors decreases when the signal-to-noise-ratio increases. 

Third, the uncertainty of the estimated factors is independent of the particular data considered, provided $\mathbf{Y}_t$ follows the DFM in \eqref{eq:DFM}.

Fourth, the MSE of the estimated factors depends on $\mathbf{\Lambda}$ and $\mathbf{\Sigma}_{\varepsilon}$ and, consequently, on the cross-sectional dimension $N$. According to assumption 1(c), when $N\to\infty$, $\mathbf{\Lambda}^{\prime} \mathbf{\Sigma}_{\varepsilon} ^{-1} \mathbf{\Lambda}$, diverges as $O(N)$. Therefore, the MSE of $\mathbf{f}^{GLS}_t$ converges to a matrix of zeros as $O(N^{-1})$, providing a proof that the GLS is $\sqrt N$--consistent by Chebychev's inequality. 
Furthermore, it follows also that, as $N\rightarrow\infty$,  the asymptotic distribution of the GLS estimator of the factors is given by
\begin{equation}
\label{eq:Asym_GLS}
\sqrt{N} \left( \mathbf{f}^{GLS}_t - \mathbf{F}_t \right) \to_d N \left( \mathbf 0_r, \mathbf{\Sigma}_{\Lambda}^{*-1} \right),
\end{equation}
where $\mathbf{\Sigma}_{\Lambda}^*$ is defined  in assumption 1.c. This result can be derived also from the  standard GLS  theory for linear regression models. Note that the MSE in \eqref{eq:MSE_1} multiplied by $N$ is an estimator of the asymptotic covariance in \eqref{eq:Asym_GLS}. This is because we can easily see that the GLS is unbiased, indeed, under \eqref{eq:DFM},
\begin{align}\label{eq:biasLS}
E(\mathbf f_t^{GLS}) = \left( \mathbf{\Lambda}^{\prime} \mathbf{\Sigma}^{-1}_{\varepsilon} \mathbf{\Lambda} \right) ^{-1} \mathbf{\Lambda}^{\prime} \mathbf{\Sigma}^{-1}_{\varepsilon} \left(\mathbf \Lambda \mathbf F_t + E(\boldsymbol {\varepsilon}_t)\right) = \mathbf F_t.
\end{align}

%

Given that, as mentioned above, computing the GLS is problematic due to the required inversion of $\mathbf{\Sigma}_{\varepsilon}$, one can, alternatively, estimate the factors by Weighted Least Squares (WLS) treating $\mathbf{\Sigma}_{\varepsilon}$ as if it were diagonal, i.e. by neglecting the cross-sectional correlations of the idiosyncratic components, but still allowing for cross-sectional heteroscedasticity, as follows (see Bartlett, 1937, Lawley and Maxwell, 1971, and Bai and Li,  2016)
\begin{equation}
\label{eq:WLS}
\mathbf{f}^{WLS}_t=\left( \mathbf{\Lambda}^{\prime} \mathbf{\Sigma}^{*-1}_{\varepsilon} \mathbf{\Lambda} \right)^{-1} \mathbf{\Lambda}^{\prime} \mathbf{\Sigma}^{*-1}_{\varepsilon} \mathbf{Y}_t,
\end{equation}
where $\mathbf{\Sigma}^{*}_{\varepsilon}=\mathrm{diag}(\mathbf{\Sigma}_{\varepsilon})$ as defined in assumption 1(e). Note that if we compute the WLS in \eqref{eq:WLS} using the PC estimator of the loadings and a consistent estimator of the idiosyncratic variances, then we would obtain the WPC estimator studied by Breitung and Tenhofen (2011).

From \eqref{eq:WLS} we can directly derive the finite sample MSE of $\mathbf{f}^{WLS}_t$:
\begin{align}
\label{eq:MSE_3}
\text{MSE}(\mathbf{f}^{WLS}_t)
&=\left( \mathbf{\Lambda}^{\prime} \mathbf{\Sigma}^{*-1}_{\varepsilon} \mathbf{\Lambda} \right)^{-1} 
\mathbf{\Lambda}^{\prime} \mathbf\Sigma^{*-1}_{\varepsilon} \mathbf{\Sigma}_{\varepsilon} \mathbf{\Sigma}^{*-1}_{\varepsilon} \mathbf{\Lambda} 
\left( \mathbf{\Lambda}^{\prime} \mathbf{\Sigma}^{*-1}_{\varepsilon} \mathbf{\Lambda} \right)^{-1}.
\end{align}
It is important to note that now the MSE matrix of $\mathbf{f}_t^{WLS}$ is not diagonal unless the idiosyncratic components are uncorrelated so that $\mathbf{\Sigma}_{\varepsilon}$ is truly diagonal, i.e. $\mathbf{\Sigma}_{\varepsilon}=\mathbf{\Sigma}_{\varepsilon}^*$. In this latter case,
the MSE of $\mathbf{f}^{WLS}_t$ becomes 
\begin{equation}
\label{eq:MSE_3_0}
\text {MSE}^{(0)}(\mathbf{f}^{WLS}_t)=\left( \mathbf{\Lambda}^{\prime} \mathbf{\Sigma}^{*-1}_{\varepsilon} \mathbf{\Lambda} \right)^{-1},
\end{equation} 
which is the Gauss-Markov lower bound for heteroscedastic errors. This is the case originally considered by Bartlett (1937) (see also Bai and Li, 2012). The difference between the MSE of WLS factors in (\ref{eq:MSE_3}) and the MSE computed as in (\ref{eq:MSE_3_0}) when $\mathbf{\Sigma}_{\varepsilon}$ is full, is analysed by Fresoli, Poncela and Ruiz (2024), who show that it can be either positive or negative definite depending on the signs and magnitudes of the idiosyncratic cross-correlations and variances.

Now, according to assumptions 1(e) and 1(f), when $N\to\infty$, both $\mathbf{\Lambda}^{\prime} \mathbf{\Sigma}^{*-1}_ {\varepsilon} \mathbf{\Lambda}$ and $\mathbf{\Lambda}^{\prime} \mathbf{\Sigma}_{\varepsilon}^{*-1} \mathbf{\Sigma}_{\varepsilon} \mathbf{\Sigma}_{\varepsilon}^{*-1}\mathbf{\Lambda}$ diverge as $O(N)$. Consequently, the MSE of $\mathbf{f}_t^{WLS}$ converges to a matrix of zeros as $O(N^{-1})$. It follows that, as $N\to\infty$,
the WLS estimator of the factors in (\ref{eq:WLS}) is $\sqrt N$--consistent by Chebychev's inequality, and has the following asymptotic distribution
\begin{equation}
\label{eq:Asym_WLS}
\sqrt{N} \left( \mathbf{f}^{WLS}_t - \mathbf{F}_t \right) \to_d N\left( \mathbf 0_r, \mathbf{\Sigma}^{**-1}_{\Lambda} \mathbf{\Sigma}^{\ddag}_{\Lambda}
 \mathbf{\Sigma}^{**-1}_{\Lambda}\right),
\end{equation}
where  $\mathbf{\Sigma}^{**}_{\Lambda}$ and $\mathbf{\Sigma}^{\ddag}_{\Lambda}$ are defined in assumptions 1(e) and 1(f), respectively. The sandwich form of the asymptotic covariance matrix reflects the mis-specification due to considering only the diagonal elements of the idiosyncratic covariance. Again, note that the MSE in \eqref{eq:MSE_3} multiplied by $N$ is an estimator of the asymptotic covariance in \eqref{eq:Asym_WLS}. This is because, using the same argument in \eqref{eq:biasLS}, we can easily see that the WLS is unbiased.

Finally, when $\mathbf \Sigma_{\varepsilon}$ is treated as if it were spherical, i.e. it is replaced with $\sigma^2_{\varepsilon}\mathbf I_N$, for some scalar $\sigma_{\varepsilon}^2>0$, then the GLS in \eqref{eq:GLS} would boil down to the following Ordinary Least Squares (OLS) estimator 
\begin{equation}
\label{eq:LS}
\mathbf{f}^{OLS}_t=\left( \mathbf{\Lambda}^{\prime} \mathbf{\Lambda} \right) ^{-1} \mathbf{\Lambda}^{\prime} \mathbf{Y}_t.
\end{equation}
Note that if we computed the OLS in \eqref{eq:LS} using the PC estimator of the loadings, then we would obtain the usual PC estimator.

From \eqref{eq:LS} we can directly derive finite sample MSE of $\mathbf{f}_t^{OLS}$:
\begin{equation}
\label{eq:MSE_2}
\text{MSE}(\mathbf{f}^{OLS}_t)
=(\mathbf{\Lambda}^{\prime} \mathbf{\Lambda})^{-1} \mathbf{\Lambda}^{\prime} \mathbf{\Sigma}_{\varepsilon} \mathbf{\Lambda} ( \mathbf{\Lambda}^{\prime} \mathbf{\Lambda})^{-1},
\end{equation}
which coincides with the finite sample MSE of the PC estimator of the factors when the  identifying restrictions, $\frac1 T \mathbf{F}^{\prime}\mathbf{F}=I_r$ and $\frac 1 N\mathbf{\Lambda}^{\prime}\mathbf{\Lambda}$ diagonal, are satisfied for all $N,T$ (see Bai and Ng, 2013, and Appendix A for the equivalence between the MSEs). 
In general, the MSE matrix of the OLS is not diagonal due to the presence of idiosyncratic cross-correlations, i.e., since $\mathbf{\Sigma}_{\varepsilon}$ is a full matrix.
 
Note that if $\mathbf \Sigma_{\varepsilon}=\sigma^2_\varepsilon\mathbf I_N$ so that the idiosyncratic components were truly spherical then the MSE of $\mathbf{f}^{OLS}_t$ would be
\begin{equation}
\label{eq:MSE_2_0}
\text{MSE}^{(0)}(\mathbf{f}^{OLS}_t) = \sigma^2_\varepsilon(\mathbf{\Lambda}^{\prime} \mathbf{\Lambda})^{-1},
\end{equation}
which is the Gauss-Markov lower bound for cross-sectionally homoscedastic and uncorrelated errors. The difference between the MSE of the OLS in \eqref{eq:MSE_2} and \eqref{eq:MSE_2_0}, is analysed in Barigozzi and Luciani (2024) who show that WLS/WPC is always more efficient than OLS/PC if $\mathbf{\Sigma}_{\varepsilon}$ is sufficiently sparse (this is obvious if $\mathbf{\Sigma}_{\varepsilon}$ is diagonal), i.e.,
$\text{MSE}(\mathbf{f}^{OLS}_t)-\text{MSE}(\mathbf{f}^{WLS}_t)$ is positive definite.

Finally, as implied by assumptions 1(a) and 1(d), when $N\to\infty$, the matrices $\mathbf{\Lambda}^{\prime} \mathbf{\Lambda}$ and $\mathbf{\Lambda}^{\prime} \mathbf{\Sigma}_{\varepsilon} \mathbf{\Lambda}$ diverge as $O(N)$. Therefore, the MSE of $\mathbf{f}_t^{OLS}$ converges to a matrix of zeros as $O(N^{-1})$.
This implies that, as $N\to\infty$,
the OLS estimator of the factors in (\ref{eq:LS}) is $\sqrt N$--consistent by Chebychev's inequality, and has the following asymptotic distribution
\begin{equation}
\label{eq:Asym_LS}
\sqrt{N} \left( \mathbf{f}^{OLS}_t - \mathbf{F}_t \right) \to_d N\left( \mathbf 0_r,\mathbf{\Sigma}_{\Lambda} ^{-1} \mathbf{\Sigma}^{\dagger}_{\Lambda}\mathbf {\Sigma}_{\Lambda} ^{-1}\right),
\end{equation}
where $\mathbf{\Sigma}^{\dagger}_{\Lambda}$ is defined in assumption 1.d. As before, note that the MSE in \eqref{eq:MSE_2} multiplied by $N$ is an estimator of the asymptotic covariance in \eqref{eq:Asym_LS}.
This is because, using the same argument in \eqref{eq:biasLS}, we can easily see that the OLS is unbiased.

Table \ref{tab:Summary} summarizes the finite sample MSEs of the factors described in this and the next sections.

\begin{table}[h!]
\centering
\resizebox{0.95\textwidth}{!}
{
\begin{tabular}{|c|c|c|c|}
\hline
\hline
&\multicolumn{3}{c}{True idiosyncratic covariance}\\
\hline 
& $\mathbf{\Sigma}_{\varepsilon}=\sigma^2_{\varepsilon} \mathbf{I}_{N}$ & $\mathbf{\Sigma}_{\varepsilon}=\mathbf{\Sigma}_{\varepsilon}^* $ & $\mathbf{\Sigma}_{\varepsilon}$ \\
&  Spherical&  Diagonal & Full\\
\hline
\hline
\textbf{GLS} & $\sigma^2_{\varepsilon} (\mathbf{\Lambda}^{\prime} \mathbf{\Lambda} )^{-1}$  &  $\mathbf{\Omega}^{*-1}$ &  $\mathbf{\Omega}^{-1}$  \\
&&&\\
\textbf{WLS} & $\sigma^2_{\varepsilon} (\mathbf{\Lambda}^{\prime} \mathbf{\Lambda} )^{-1}$ & $\mathbf{\Omega}^{*-1}$ &  $\mathbf{\Omega}^{*-1} \mathbf{\Lambda}^{\prime} \mathbf{\Sigma}_{\varepsilon}^{*-1} \mathbf{\Sigma}_{\varepsilon}  \mathbf{\Sigma}_{\varepsilon}^{*-1} \mathbf{\Lambda} \mathbf{\Omega}^{*-1}$ \\
&&&\\
\textbf{OLS} & $\sigma^2_{\varepsilon} (\mathbf{\Lambda}^{\prime} \mathbf{\Lambda} )^{-1}$ & $(\mathbf{\Lambda}^{\prime} \mathbf{\Lambda} )^{-1} \mathbf{\Lambda}^{\prime} \mathbf{\Sigma}_{\varepsilon}^* \mathbf{\Lambda} (\mathbf{\Lambda}^{\prime} \mathbf{\Lambda} )^{-1}$ & $(\mathbf{\Lambda}^{\prime} \mathbf{\Lambda} )^{-1} \mathbf{\Lambda}^{\prime} \mathbf{\Sigma}_{\varepsilon} \mathbf{\Lambda} (\mathbf{\Lambda}^{\prime} \mathbf{\Lambda} )^{-1}$ \\
\hline
\hline
\textbf{fLP} & $\sigma^2_{\varepsilon}(\sigma^2_{\varepsilon} \mathbf{I}_r + \mathbf{\Lambda}^{\prime} \mathbf{\Lambda})^{-1}$ &  $(\mathbf{I}_r+ \mathbf{\Omega}^*)^{-1}$ & $(\mathbf{I}_r+ \mathbf{\Omega})^{-1}$ \\
&&&\\
\textbf{dLP} &  $\sigma^2_{\varepsilon}(\sigma^2_{\varepsilon} \mathbf{I}_r + \mathbf{\Lambda}^{\prime} \mathbf{\Lambda})^{-1}$  &  $ \left(\mathbf{I}_r + \mathbf{\Omega}^* \right)^{-1}$ & $\mathbf{I}_r + \mathbf{C}$ \\
&&&\\
\textbf{sLP} &  $\sigma^2_{\varepsilon}(\sigma^2_{\varepsilon} \mathbf{I}_r + \mathbf{\Lambda}^{\prime} \mathbf{\Lambda})^{-1}$  & $\mathbf{I}_r+ \mathbf{B}$ & $\mathbf{I}_r + \mathbf{A}$  \\
\hline
\hline
\textbf{fKF} & $ \sigma^2_{\varepsilon} \left( \mathbf{\Lambda}^{\prime} \mathbf{\Lambda} \right)^{-1}(\sigma^2_{\varepsilon} \mathbf{I}_r + \mathbf{\Lambda}^{\prime} \mathbf{\Lambda D})^{-1} \mathbf{\Lambda}^{\prime} \mathbf{\Lambda D}$ & $ \mathbf{\Omega}^{*-1}\left(\mathbf{ I}_r+ \mathbf{\Omega}^* \mathbf{D} \right)^{-1} \mathbf{\Omega}^* \mathbf{D}$ & 
$\mathbf{\Omega}^{-1} \left( \mathbf{I}_r+ \mathbf{\Omega D} \right)^{-1} \mathbf{\Omega D}$ \\
&&&\\
\textbf{dKF} & $ \sigma^2_{\varepsilon} \left( \mathbf{\Lambda}^{\prime} \mathbf{\Lambda} \right)^{-1}(\sigma^2_{\varepsilon} \mathbf{I}_r + \mathbf{\Lambda}^{\prime} \mathbf{\Lambda D})^{-1} \mathbf{\Lambda}^{\prime} \mathbf{\Lambda D}$ & $\mathbf{\Omega}^{*-1} \left( \mathbf{I}_r + \mathbf{\Omega}^*\mathbf D \right)^{-1} \mathbf{\Omega}^* \mathbf D$ &See expression (\ref{eq:MSE_wrongKF_2}) \\
&&&\\
\textbf{sKF} & $ \sigma^2_{\varepsilon} \left( \mathbf{\Lambda}^{\prime} \mathbf{\Lambda} \right)^{-1} \left( \sigma^2_{\varepsilon} \mathbf{I}_r + \mathbf{\Lambda}^{\prime} \mathbf{\Lambda} \mathbf {D} \right)^{-1} \mathbf{\Lambda}^{\prime} \mathbf{\Lambda} \mathbf{D}$ & See expression (\ref{eq:MSE_wrongKF_3}) with $\mathbf{\Sigma}_{\varepsilon}$ substituted by $\mathbf{\Sigma}^*_{\varepsilon}$  & See expression (\ref{eq:MSE_wrongKF_3}) \\
\hline
\end{tabular}
}

\caption{Summary results on MSE of factors estimated with known parameters. The formulas for the KF estimators are those obtained when reaching convergence to the steady-state.\\
Notation:\\
$\mathbf{\Omega}^*= \mathbf{\Lambda}^{\prime} \mathbf{\Sigma}^{*-1}_{\varepsilon} \mathbf{\Lambda}$; $\mathbf{\Omega}= \mathbf{\Lambda}^{\prime} \mathbf{\Sigma}^{-1}_{\varepsilon} \mathbf{\Lambda}$;\\
$\mathbf{A}=\mathbf{\Lambda}^{\prime} \left( \mathbf{\Lambda \Lambda}^{\prime} + \sigma^2_{\varepsilon} \mathbf{I}_N \right)^{-1} \left( \mathbf{\Lambda \Lambda}^{\prime} + \mathbf{\Sigma}_{\varepsilon}\right)\left( \mathbf{\Lambda \Lambda}^{\prime} + \sigma^2_{\varepsilon} \mathbf{I}_N \right)^{-1} \mathbf{\Lambda} - 2 \mathbf{\Lambda}^{\prime} \left( \mathbf{\Lambda \Lambda}^{\prime} + \sigma^2_{\varepsilon} \mathbf{I}_N \right)^{-1} \mathbf{\Lambda}$;\\ $\mathbf{B}=\mathbf{\Lambda}^{\prime} \left( \mathbf{\Lambda \Lambda}^{\prime} + \sigma^2_{\varepsilon} \mathbf{I}_N \right)^{-1} \left( \mathbf{\Lambda \Lambda}^{\prime} + \mathbf{\Sigma}^*_{\varepsilon}\right)\left( \mathbf{\Lambda \Lambda}^{\prime} + \sigma^2_{\varepsilon} \mathbf{I}_N \right)^{-1} \mathbf{\Lambda} - 2 \mathbf{\Lambda}^{\prime} \left( \mathbf{\Lambda \Lambda}^{\prime} + \sigma^2_{\varepsilon} \mathbf{I}_N \right)^{-1} \mathbf{\Lambda}$;\\
 $\mathbf{C}=\mathbf{\Lambda}^{\prime} \left(\mathbf{\Lambda \Lambda}^{\prime} + \mathbf{\Sigma}^*_{\varepsilon} \right)^{-1} \left(\mathbf{\Lambda \Lambda}^{\prime} + \mathbf{\Sigma}_{\varepsilon} \right) \left(\mathbf{\Lambda \Lambda}^{\prime} + \mathbf{\Sigma}^*_{\varepsilon} \right)^{-1} \mathbf{\Lambda} - 2 \mathbf{\Lambda}^{\prime} \left(\mathbf{\Lambda \Lambda}^{\prime} + \mathbf{\Sigma}^*_{\varepsilon} \right)^{-1} \mathbf{\Lambda}$; \\
$\mathbf{D}= \mathbf{\Phi} \bar{\mathbf{P}} \mathbf{\Phi}^{\prime} + \mathbf{\Sigma}_{\eta}$. 
}
\label{tab:Summary}
\end{table}

\section{ Linear Projections and Kalman Filter}
\label{sec:KFS}

The Least Squares estimators described above, rely on treating the factors as if they were fixed parameters, i.e., deterministic. In this section, we derive the MSEs of the factors when they are considered as random variables. We consider first the case in which they are assumed to be \textit{iid} and, second, when they are assumed to be serially correlated. 

\subsection{Serially independent factors: Linear Projections}

If the factors are \textit{iid} random variables, the optimal (in mean-squared sense) estimator of the factors at each time period, $t$, is given by the (full) Linear Projection (fLP) of $\mathbf{F}_t$ on $\mathbf{Y}_t$, and  given by (see Thomson, 1951, Lawley and Maxwell, 1971, Bai and Li, 2012, and Mao \textit{et al}., 2024)\footnote{The expressions of the linear projection and its associated MSE can be obtained as a particular case of the Kalman filter as derived by Kalman (1960). 
Note that under normality, the linear projection, $\mathbf{f}_t^{fLP}$, can be interpreted as the conditional mean of $\mathbf{F}_t$ given $\mathbf{Y}_t$.}
\begin{align}
\label{eq:LP_1}
\mathbf{f}_t^{fLP}= \mathrm{Proj}\left\lbrace \mathbf{F}_t | \mathbf{Y}_t \right\rbrace &
= \mathbf{\Lambda}^{\prime} \left( \mathbf{\Lambda} \mathbf{\Sigma}_F \mathbf{\Lambda}^{\prime} + \mathbf{\Sigma}_{\varepsilon} \right) ^{-1} \mathbf{Y}_t\nonumber\\
&= \mathbf{\Lambda}^{\prime} \left( \mathbf{\Lambda}  \mathbf{\Lambda}^{\prime} + \mathbf{\Sigma}_{\varepsilon} \right) ^{-1} \mathbf{Y}_t,
\end{align}
where in the second line we use the fact that when $\frac 1T\mathbf F^\prime\mathbf F=\mathbf I_r$, for all $T$, we then also have $\mathbf{\Sigma}_F=\mathbf I_r$ because of assumption 2(c). 

Now, using Lemma 3.3 of Duncan and Horn (1972), it is possible to show that\footnote{Lemma 3.3 of Duncan and Horn (1972) states that, if $\mathbf{S}=\left(\mathbf{M}^{-1} + \mathbf{X}^{\prime} \mathbf{R} ^{-1} \mathbf{X} \right) ^{-1}$, then $\mathbf{S}=\mathbf{M}-\mathbf{M X}^{\prime} \left( \mathbf{R} + \mathbf{X M X}^{\prime}\right) ^{-1} \mathbf{X M}$.} 
\begin{equation}
\label{eq:DH}
(\mathbf{\Lambda} \mathbf{\Lambda}^{\prime} + \mathbf{\Sigma}_{\varepsilon})^{-1}= \mathbf{\Sigma}_{\varepsilon}^{-1} - \mathbf{\Sigma}_{\varepsilon}^{-1} \mathbf{\Lambda} (\mathbf{\Lambda}^{\prime} \mathbf{\Sigma}_{\varepsilon} ^{-1} \mathbf{\Lambda}+\mathbf{I}_r)^{-1} \mathbf{\Lambda}^{\prime} \mathbf{\Sigma}_{\varepsilon}^{-1}.
\end{equation}
Then, using the Woodbury formula, we can see that\footnote{The Woodbury formula establishes that $\left(\mathbf{A}+\mathbf{UCV} \right) ^{-1}= \mathbf{A}^{-1} - \mathbf{A}^{-1} \mathbf{U} \left( \mathbf{C}^{-1}+ \mathbf{VA}^{-1}\mathbf{U} \right) ^{-1} \mathbf{V A}^{-1}$, and, in particular, $\left(\mathbf{A}+\mathbf{B} \right) ^{-1}=\mathbf{A}^{-1} - \mathbf{A}^{-1} \left(  \mathbf{I}+\mathbf{A B}^{-1}\right)^{-1}$.}
\begin{equation}
\label{eq:W}
(\mathbf{\Lambda}^{\prime} \mathbf{\Sigma}_{\varepsilon} ^{-1} \mathbf{\Lambda}+\mathbf{I}_r)^{-1}=\left(\mathbf{\Lambda}^{\prime} \mathbf{\Sigma}_{\varepsilon} ^{-1} \mathbf{\Lambda}\right)^{-1} - \left(\mathbf{\Lambda}^{\prime} \mathbf{\Sigma}_{\varepsilon} ^{-1} \mathbf{\Lambda}\right)^{-1} \left(\mathbf{\Lambda}^{\prime} \mathbf{\Sigma}_{\varepsilon} ^{-1} \mathbf{\Lambda} + \mathbf{I}_r\right)^{-1}.
\end{equation}
Therefore, substituting (\ref{eq:DH}) and (\ref{eq:W}) into (\ref{eq:LP_1}), we obtain
\begin{align}
\label{eq:LP_2}
\mathbf{f}_t^{fLP}&= \mathbf{\Lambda}^{\prime} \mathbf{\Sigma}_{\varepsilon}^{-1} \mathbf{Y}_t - \mathbf{\Lambda}^{\prime} \mathbf{\Sigma}_{\varepsilon} ^{-1} \mathbf{\Lambda} \left( \mathbf{\Lambda}^{\prime} \mathbf{\Sigma}_{\varepsilon} ^{-1} \mathbf{\Lambda}+\mathbf{I}_r\right)^{-1} \mathbf{\Lambda}^{\prime} \mathbf{\Sigma}_{\varepsilon}^{-1} \mathbf{Y}_t \nonumber\\
&= \mathbf{\Lambda}^{\prime} \mathbf{\Sigma}_{\varepsilon}^{-1} \mathbf{Y}_t - \mathbf{\Lambda}^{\prime} \mathbf{\Sigma}_{\varepsilon}^{-1} \mathbf{Y}_t + \left(\mathbf{\Lambda}^{\prime} \mathbf{\Sigma}_{\varepsilon} ^{-1} \mathbf{\Lambda} + \mathbf{I}_r\right)^{-1} \mathbf{\Lambda}^{\prime} \mathbf{\Sigma}_{\varepsilon}^{-1} \mathbf{Y}_t\nonumber\\
& =\left( \mathbf{\Lambda}^{\prime} \mathbf{\Sigma}_{\varepsilon} ^{-1} \mathbf{\Lambda}+ \mathbf{I}_r \right) ^{-1} \mathbf{\Lambda}^{\prime} \mathbf{\Sigma}_{\varepsilon}^{-1} \mathbf{Y}_t.
\end{align}
It can be proved that for the fLP in (\ref{eq:LP_2}) the MSE conditional on a realization of the data $\mathbf Y_t$ is given by
\begin{align}
\label{eq:MSE_LP}
E\left[ \left(\mathbf{f}_t^{fLP}-\mathbf{F}_t\right) \left(\mathbf{f}_t^{fLP}-\mathbf{F}_t\right)^{\prime} |\mathbf{Y}_t \right] 
&= \mathbf{I}_r - \mathbf{\Lambda}^{\prime} (\mathbf{\Lambda \Lambda}^{\prime} + \mathbf{\Sigma}_{\varepsilon})^{-1} \mathbf{\Lambda} \nonumber\\
&= \mathbf{I}_r - \mathbf{\Lambda}^{\prime} \mathbf{\Sigma}_{\varepsilon} ^{-1} \mathbf{\Lambda} + \mathbf{\Lambda}^{\prime} \mathbf{\Sigma}_{\varepsilon} ^{-1} \mathbf{\Lambda} \left(\mathbf{I}_r+\mathbf{\Lambda}^{\prime} \mathbf{\Sigma}_{\varepsilon} ^{-1} \mathbf{\Lambda}\right)^{-1} \mathbf{\Lambda}^{\prime} \mathbf{\Sigma}_{\varepsilon} ^{-1} \mathbf{\Lambda}\nonumber\\ 
&= \mathbf{I}_r-\left(\mathbf{I}_r+\mathbf{\Lambda}^{\prime} \mathbf{\Sigma}_{\varepsilon} ^{-1} \mathbf{\Lambda}\right)^{-1} \mathbf{\Lambda}^{\prime} \mathbf{\Sigma}_{\varepsilon} ^{-1} \mathbf{\Lambda}\\ 
&= \left(\mathbf{\Lambda}^{\prime} \mathbf{\Sigma}_{\varepsilon} ^{-1} \mathbf{\Lambda}\right)^{-1} \left(\mathbf{I}_r+ \mathbf{\Lambda}^{\prime} \mathbf{\Sigma}_{\varepsilon} ^{-1} \mathbf{\Lambda}\right)^{-1} \mathbf{\Lambda}^{\prime} \mathbf{\Sigma}_{\varepsilon} ^{-1} \mathbf{\Lambda}. \nonumber
\end{align}
Equation (\ref{eq:MSE_LP}) follows from standard results on linear projections in $L^2$ (see, e.g., Kalman, 1960, and Uhlmannand Julier, 2024) and the use of \eqref{eq:DH} and \eqref{eq:W}. In the non-Gaussian case, it holds because $\mathbf{f}^{fLP}_t$, which is the best linear predictor of $\mathbf{F}_t$ given $\mathbf Y_t$, satisfies $E\left[ \left( \mathbf{F}_t - \mathbf{f}^{fLP}_t \right) \mathbf{Y}^{\prime}_t\right]=\mathbf 0$.  Whereas under Gaussianity, when $\mathbf{f}^{fLP}_t$ is the conditional mean, (\ref{eq:MSE_LP}) coincides with the variance of $\mathbf{F}_t$ conditional on $\mathbf Y_t$.

Now since the conditional MSE in (\ref{eq:MSE_LP}) does not depend on the observations, $\mathbf Y_t$, it is then equal to the unconditional MSE (by the Law of Iterated Expectations). And, since under our identifying assumptions, $\mathbf{\Lambda}^{\prime} \mathbf{\Sigma}_{\varepsilon} ^{-1} \mathbf{\Lambda}$ is diagonal, the unconditional MSE of $\mathbf f_t^{fLP}$  is then given by
\begin{equation}
\label{eq:MSE_LP_2}
\text{MSE}(\mathbf{f}_t^{fLP})
=(\mathbf{\Lambda}^{\prime} \mathbf{\Sigma}_{\varepsilon} ^{-1} \mathbf{\Lambda}+\mathbf{I}_r)^{-1},
\end{equation}
which is also diagonal. Note that the MSE in (\ref{eq:MSE_LP_2}) differs from the one derived by Mao \textit{et al}. (2024) because here we are using different identification restrictions. 

We stress that the difference between the MSE of the fLP estimator  of the factors in \eqref{eq:MSE_LP_2} and that of the GLS estimator in \eqref{eq:MSE_1}, is negative semidefinite  since, under our identifying restrictions, $(\mathbf{\Lambda}^{\prime} \mathbf{\Sigma}_{\varepsilon} ^{-1} \mathbf{\Lambda}+\mathbf{I}_r)^{-1}-(\mathbf{\Lambda}^{\prime} \mathbf{\Sigma}_{\varepsilon} ^{-1} \mathbf{\Lambda})^{-1}$ is a diagonal matrix with diagonal elements which are obviously smaller than zero. This difference between the MSEs of $\mathbf{f}_t^{fLP}$ and $\mathbf{f}_t^{GLS}$ is due to the fact that the former treats $\mathbf F_t$ as random and estimates its conditional mean, while the latter is estimating a realization of $\mathbf F_t$ and, consequently, is larger. 

Finally, as implied by assumption 1(c), when $N\to\infty$, $\mathbf{\Lambda}^{\prime} \mathbf{\Sigma}_{\varepsilon} ^{-1} \mathbf{\Lambda}$ diverges as $O(N)$, and, consequently, the MSE of $\mathbf{f}_t^{fLP}$ tends to a matrix of zeros as $O(N^{-1})$. Moreover, $(\mathbf{\Lambda}^{\prime} \mathbf{\Sigma}_{\varepsilon} ^{-1}\mathbf \Lambda+\mathbf{I}_r)^{-1}$ tends to $(\mathbf{\Lambda}^{\prime} \mathbf{\Sigma}_{\varepsilon} ^{-1} \mathbf{\Lambda})^{-1}$ with an error $O(N^{-1})$. Therefore, the fLP and GLS estimators coincide asymptotically, as $N\to\infty$, with an error $O(N^{-1})$ and so their MSEs (multiplied by $N$) also coincide asymptotically. Therefore, $\mathbf{f}_t^{fLP}$ is $\sqrt N$--consistent, by Chebychev's inequality and it has the same asymptotic distribution as the one derived for the GLS in \eqref{eq:Asym_GLS}.\footnote{Note that when the loadings are estimated via PC and we employ a consistent estimator of the idiosyncratic covariance, the fLP estimator of the factors is asymptotically equivalent to the GPC estimator.} However, it is important to notice that, since now we treat factors as random, the limiting location, being $\mathbf F_t$, is random. In this case, the previous asymptotic statements should then be interpreted as conditional on a realization of the factors $\mathbf F_1,\ldots, \mathbf F_T$.
 
Moreover,  
\begin{align}\label{eq:bias}
 E[\mathbf f_t^{fLP}|\mathbf F_t] 
&=\left( \mathbf{\Lambda}^{\prime} \mathbf{\Sigma}_{\varepsilon} ^{-1} \mathbf{\Lambda}+ \mathbf{I}_r \right) ^{-1} \mathbf{\Lambda}^{\prime} \mathbf{\Sigma}_{\varepsilon}^{-1} \mathbf\Lambda\mathbf F_t
+\left( \mathbf{\Lambda}^{\prime} \mathbf{\Sigma}_{\varepsilon} ^{-1} \mathbf{\Lambda}+ \mathbf{I}_r \right) ^{-1} \mathbf{\Lambda}^{\prime} \mathbf{\Sigma}_{\varepsilon}^{-1}E[\boldsymbol{\varepsilon}_t|\mathbf F_t]
\nonumber\\
 &= \mathbf F_t + O\left(\frac 1N\right)+ \left( \mathbf{\Lambda}^{\prime} \mathbf{\Sigma}^{-1}_{\varepsilon} \mathbf{\Lambda} \right) ^{-1} \mathbf{\Lambda}^{\prime} \mathbf{\Sigma}^{-1}_{\varepsilon}  E[\boldsymbol{\varepsilon}_t|\mathbf F_t],
 \end{align}
 so if $E[\boldsymbol{\varepsilon}_t|\mathbf F_t]=\mathbf 0$, then the fLP is asymptotically unbiased and the MSE in \eqref{eq:MSE_LP_2} multiplied by $N$ is still a consistent estimator of the asymptotic variance in \eqref{eq:Asym_GLS}. Sufficient conditions would be that either $\boldsymbol{\varepsilon}_t$ and $\mathbf F_t$ are independent or $\mathbf Y_t$ and $\mathbf F_t$ are jointly normal.

Consider now the LP obtained by wrongly treating the idiosyncratic covariance matrix as if it were is diagonal, so that  in \eqref{eq:LP_1} and \color{black} \eqref{eq:LP_2} we replace $\mathbf{\Sigma}_{\varepsilon}$ with $\mathbf{\Sigma}_{\varepsilon}^*=\text{diag}(\mathbf{\Sigma}_{\varepsilon})$. Then, the linear projection is obtained as follows
\begin{equation}
\label{eq:LPD}
\mathbf{f}_t^{dLP}=\mathbf{\Lambda}^{ \prime} \left( \mathbf{\Lambda \Lambda}^{\prime} + \mathbf{\Sigma}^*_{\varepsilon} \right)^{-1} \mathbf{Y}_t = \left(  \mathbf{\Lambda}^{\prime} \mathbf{\Sigma}^{*-1}_{\varepsilon} \mathbf{\Lambda}  + \mathbf{I}_r \right)^{-1} \mathbf{\Lambda}^{\prime} \mathbf{\Sigma}^*_{\varepsilon} \mathbf{Y}_t.
\end{equation}
Now, if $\mathbf{\Sigma}_{\varepsilon}$ were truly diagonal then the MSE of $\mathbf{f}_t^{dLP}$ would be (see \eqref{eq:MSE_LP})
\begin{equation}
\label{eq:MSE_LPD}
\text{MSE}^{(0)}(\mathbf{f}_t^{dLP})= \mathbf I_r-\mathbf{\Lambda}^{\prime} \left(\mathbf{\Lambda \Lambda}^{\prime} + \mathbf{\Sigma}^*_{\varepsilon} \right)^{-1} \mathbf{\Lambda}=
\left(  \mathbf{\Lambda}^{\prime} \mathbf{\Sigma}^{*-1}_{\varepsilon} \mathbf{\Lambda}  +\mathbf{I}_r \right)^{-1},
\end{equation}
by Lemma 3.3 of Duncan and Horn (1972).
But in general, since $\mathbf{\Sigma}_{\varepsilon}$ it is not diagonal, the MSE of $\mathbf{f}_t^{dLP}$ is given by 
\begin{equation}
\begin{split}
\label{eq:MSE_DLP_2}
\text{MSE}(\mathbf{f}_t^{dLP})=&\,\mathbf{I}_r + \mathbf{\Lambda}^{\prime} \left(\mathbf{\Lambda \Lambda}^{\prime} + \mathbf{\Sigma}^*_{\varepsilon} \right)^{-1} \left(\mathbf{\Lambda \Lambda}^{\prime} + \mathbf{\Sigma}_{\varepsilon} \right) \left(\mathbf{\Lambda \Lambda}^{\prime} + \mathbf{\Sigma}^*_{\varepsilon} \right)^{-1} \mathbf{\Lambda}\\ 
&- 2 \mathbf{\Lambda}^{\prime} \left(\mathbf{\Lambda \Lambda}^{\prime} + \mathbf{\Sigma}^*_{\varepsilon} \right)^{-1} \mathbf{\Lambda}.
\end{split}
\end{equation}
This formula can be obtained from the analogous formula, derived by Harvey and Delle Monache (2009), for the Kalman filter (see \eqref{eq:MSE_wrongKF_2} below), but when the factors are \textit{iid}.

In Appendix A we prove two important results. First, when $N\to\infty$, the MSE in \eqref{eq:MSE_DLP_2} tends to a matrix of zeros as $O(N^{-1})$ (see \eqref{eq:MSE_limit_result}). 
Second, the dLP and WLS estimators coincide asymptotically, as $N\to\infty$, with an error $O(N^{-1})$ and so their MSEs (multiplied by $N$) also coincide asymptotically (see \eqref{eq:MSE_dLP_vs_WLS}).\footnote{Note that, when the loadings and idiosyncratic variances are estimated via PC, the dLP estimator of the factors is asymptotically equivalent to the WPC estimator.} Therefore, conditionally on a realization $\left\{ \mathbf F_1, \ldots, \mathbf F_T \right\}$, the dLP estimator is $\sqrt N$--consistent, by Chebychev's inequality, and asymptotically normal with the same distribution as the WLS estimator in \eqref{eq:Asym_WLS}. Moreover, following the similar arguments as in \eqref{eq:bias}, it is straightforward to see that, if $E[\boldsymbol{\varepsilon}_t|\mathbf F_t]=\mathbf 0$, then the dLP is asymptotically unbiased and then its MSE in  \eqref{eq:MSE_DLP_2} multiplied by $N$ is a consistent estimator of the asymptotic variance in \eqref{eq:Asym_WLS}.


Finally, consider the case in which the idiosyncratic covariance matrix is wrongly modelled as if it were spherical, that is the factors are extracted using the following linear projection based on using the misspecified matrix $\sigma^2_{\varepsilon} \mathbf{I}_N$ in place of $\mathbf{\Sigma}_{\varepsilon}$ in \eqref{eq:LP_1} and \eqref{eq:LP_2}
\begin{equation}
\label{eq:SLP}
\mathbf{f}_t^{sLP}= \mathbf{\Lambda}^{\prime} \left( \mathbf{\Lambda \Lambda}^{\prime} + \sigma^2_{\varepsilon} \mathbf{I}_N \right)^{-1}  \mathbf{Y}_t=\left( \sigma^2_{\varepsilon} \mathbf{I}_r + \mathbf{\Lambda}^{\prime} \mathbf{\Lambda} \right)^{-1} \mathbf{\Lambda}^{\prime} \mathbf{Y}_t.
\end{equation}

Now if the idiosyncratic components were truly spherical, the MSE of $\mathbf{f}_t^{sLP}$ would be (see \eqref{eq:MSE_LP})
\begin{align}
\label{eq:MSE_SLP_1}
\text{MSE}^{(0)}(\mathbf{f}_t^{sLP}) &= \mathbf{I}_r - \mathbf{\Lambda}^{\prime} \left( \mathbf{\Lambda \Lambda}^{\prime} + \sigma^2_{\varepsilon} \mathbf{I}_N \right)^{-1} \mathbf{\Lambda} \nonumber\\
&= \left( \frac{\mathbf{\Lambda}^{\prime} \mathbf{\Lambda}}{\sigma^2_{\varepsilon}}\right)^{-1} \left(\frac{\mathbf{\Lambda}^{\prime} \mathbf{\Lambda}}{\sigma^2_{\varepsilon}}+ \mathbf{I}_r \right)^{-1}\frac{\mathbf{\Lambda}^{\prime} \mathbf{\Lambda}}{\sigma^2_{\varepsilon}}\nonumber\\
&=\left(\frac{\mathbf{\Lambda}^{\prime} \mathbf{\Lambda}}{\sigma^2_{\varepsilon}}+ \mathbf{I}_r  \right)^{-1}\nonumber\\ 
&= \sigma^2_{\varepsilon} \left( \mathbf{\Lambda}^{\prime} \mathbf{\Lambda} + \sigma^2_{\varepsilon} \mathbf{I}_r\right)^{-1},
\end{align}
where we use the fact that  given the identifying restrictions, $\mathbf{\Lambda}^{\prime} \mathbf{\Lambda}$ is diagonal. In this case the MSE matrix of $\mathbf{f}_t^{sLP}$ in (\ref{eq:MSE_SLP_1}) is diagonal.

However, in general when $\mathbf{\Sigma}_{\varepsilon}$ is a full matrix then \eqref{eq:MSE_SLP_1} is  not the MSE of $\mathbf{f}_t^{sLP}$. Indeed, if $\mathbf{\Sigma}_{\varepsilon}$ is full, the correct MSE can be obtained direectly from \eqref{eq:MSE_DLP_2} when replacing $\mathbf{\Sigma}^*_\varepsilon$ with $\sigma^2_\varepsilon\mathbf I_N$, that is,
\begin{equation}
\begin{split}
\label{eq:MSE_SLP_2}
\text{MSE}(\mathbf{f}_t^{sLP}) = &\, \mathbf{I}_r+\mathbf{\Lambda}^{\prime} \left( \mathbf{\Lambda \Lambda}^{\prime} + \sigma^2_{\varepsilon} \mathbf{I}_N \right)^{-1} \left( \mathbf{\Lambda \Lambda}^{\prime} + \mathbf{\Sigma}_{\varepsilon}\right)\left( \mathbf{\Lambda \Lambda}^{\prime} + \sigma^2_{\varepsilon} \mathbf{I}_N \right)^{-1} \mathbf{\Lambda}  \\
&- 2 \mathbf{\Lambda}^{\prime} \left( \mathbf{\Lambda \Lambda}^{\prime} + \sigma^2_{\varepsilon} \mathbf{I}_N \right)^{-1} \mathbf{\Lambda},
\end{split}
\end{equation}
which is not diagonal any more.

The asymptotic behaviour of the sLP estimator under the miss-specification of a spherical idiosyncratic covariance matrix follows directly from the previous result for the diagonal mis-specification. Consequently, as $N \rightarrow \infty$, the MSE of the sLP estimator tends to a matrix of zeros as $O(N^{-1})$ (see \eqref{eq:MSE_limit_result_sLP} in Appendix A). Moreover, as $N\to\infty$, the sLP is asymptotically equivalent to the OLS estimator, with an error $O(N^{-1})$, and their MSEs (multiplied by $N$) are asymptotically equivalent too (see \eqref{eq:MSE_sLP_vs_OLS} in Appendix A). Hence, conditional on a realization $\mathbf F_1, \ldots, \mathbf F_T$, the sLP estimator is $\sqrt N$--consistent, by Chebychev's inequality.  and asymptotically normal with the same distribution as the OLS estimator in \eqref{eq:Asym_LS}.\footnote{Note that,  when the loadings are estimated via PC, the sLP estimator of the factors is asymptotically equivalent to the PC estimator.} Moreover, following the similar arguments as in \eqref{eq:bias}, it is straightforward to see that, if $E[\boldsymbol{\varepsilon}_t|\mathbf F_t]=\mathbf 0$, then the sLP is asymptotically unbiased and then its MSE in \eqref{eq:MSE_SLP_2} multiplied by $N$ is a consistent estimator of the asymptotic variance in \eqref{eq:Asym_LS}.

\subsection{Serially correlated factors: Kalman filter}

Instead of assuming that the factors are \textit{iid} random variables, in this subsection, we assume that their dynamic dependence can be represented by a particular parametric model, typically a VAR($p$) model. In this case, after writing the DFM as a state space model, the factors can be extracted using KFS algorithms. This approach for factor extraction was developed by Engle and Watson (1981), Shumway and Stoffer (1982), Watson and Engle (1983) and Quah and Sargent (1993). Let us assume for simplicity that the vector of factors, $\mathbf{F}_t$, follows the following VAR(1) model\footnote{The results in this subsection can be easily extended to specifications with more lags by simply considering the companion form of the VAR($p$).}
\begin{equation}
\label{eq:factors}
\mathbf{F}_t=\mathbf{\Phi F}_{t-1}+\boldsymbol{\eta}_t,
\end{equation}
where the $r \times r$ matrix of autoregressive parameters, $\mathbf{\Phi}$,  is assumed to satisfy the standard  causality conditions, and $\boldsymbol{\eta}_t$ is an $r \times 1$ white noise vector with covariance matrix such that 
$\text{vec} \left(\mathbf{\Sigma}_{\eta} \right)= \left( \mathbf{I}_{r^2} -\mathbf{\Phi} \otimes \mathbf{\Phi} \right)^{-1}
\text{vec} \left({\mathbf{I}_r}\right)$, with $\text{vec}\left( \mathbf{A} \right)$ 
being the vectorization of matrix $\mathbf{A}$, and $\otimes$ being the Kroneker product. Consequently, the factors are guaranteed to  have contemporaneous covariance matrix equal to $\mathbf{I}_r$.\footnote{Note that the covariance matrix of the noises, $\boldsymbol{\eta}_t$, is restricted as a result of the identifying condition, $\frac{\mathbf F^{\prime}\mathbf F}{T}=\mathbf I_r$ and assumption 2(c), which together imply that $\mathbf \Sigma_F=\mathbf I_r$.} Note that then $\mathbf{\Sigma}_\eta = \mathbf{I}_{r}-\mathbf{\Phi \Phi}^{\prime}$.

The DFM in equation (\ref{eq:DFM}) with the factors given as in (\ref{eq:factors}) is a state space model. Consequently, given that the model parameters are known, one can run the KF to obtain filtered estimates of the underlying latent factors. The KF is defined as\footnote{Filtered estimates of the factors at time $t$ are linear projections of the factors on the observations $\mathbf{Y}_1,\ldots,\mathbf{Y}_t$. Under conditional normality, the filtered estimates of the factors are the conditional means $E(\mathbf{F}_t|\mathbf{Y}_1,\ldots,\mathbf{Y}_t)$.} 
\[
\mathbf f_t^{fKF} = \text{Proj}\{\mathbf{F}_t|\mathbf{Y}_1,\ldots,\mathbf{Y}_t\}
\]
and an implicit expression is given by the following recursions
\begin{align}
\mathbf f_t^{fKF} &= \mathbf \Phi \mathbf f_{t-1}^{fKF} + \mathbf K_t \left(\mathbf Y_t- \mathbf \Lambda \mathbf \Phi\mathbf f_{t-1}^{fKF}\right),\nonumber\\
\mathbf K_t &=  \mathbf P_{t|t-1} \mathbf \Lambda^\prime  \left(\mathbf \Lambda \mathbf P_{t|t-1} \mathbf \Lambda^\prime+\mathbf \Sigma_{\varepsilon}\right)^{-1},
\label{eq:KF_matteo}\\
\mathbf P_{t|t-1} &= \mathbf{\Phi P}_{t-1} \mathbf{\Phi}^{\prime}+ \mathbf{\Sigma}_{\eta},\nonumber\\
\mathbf P_t&= \left(\mathbf{I}_r-\mathbf{K}_t \mathbf{\Lambda}\right)\mathbf{ P}_{t|t-1},\nonumber
\end{align}
where $\mathbf{K}_t$ is the Kalman gain, $\mathbf{P}_{t|t-1}$ is the one-step-ahead conditional MSE, $\mathbf{P}_{t|t-1}=E[ (\mathbf{f}_{t|t-1}^{fKF}-\mathbf{F}_t) (\mathbf{f}_{t|t-1}^{fKF}-\mathbf{F}_t)^{\prime} |\mathbf{Y}_1,\ldots, \mathbf Y_{t-1} ]$, where $\mathbf{f}^{fKF}_{t|t-1}=\mathbf{\Phi}\mathbf{f}^{fKF}_{t-1}$ is the one-step-ahead projection, 
and $\mathbf{P}_t=E[ (\mathbf{f}_t^{fKF}-\mathbf{F}_t) (\mathbf{f}_t^{fKF}-\mathbf{F}_t)^{\prime} |\mathbf{Y}_1,\ldots, \mathbf Y_t ]$ is the conditional MSE of $\mathbf{f}_t^{fKF}$. 
Note that, since $\mathbf P_t $ does not depend on $\mathbf Y_1,\ldots, \mathbf Y_t$, then it is also the unconditional MSE  (by the Law of Iterated Expectations), i.e.,  $\text{MSE}(\mathbf{f}_t^{fKF})=\mathbf P_t$.

Now, using the Woodbury formula, the Kalman gain can also be written as
\begin{equation}
\label{eq:gain_3}
\mathbf{K}_t =  \left( \mathbf{\Lambda}^{\prime} \mathbf{\Sigma}_{\varepsilon}^{-1} \mathbf{\Lambda} + \mathbf{P}_{t|t-1}^{-1} \right)^{-1} \mathbf{\Lambda}^{\prime} \mathbf{\Sigma}^{-1}_{\varepsilon},
\end{equation}
and, by substituting (\ref{eq:gain_3}) in the expression of $\mathbf{P}_t$, we obtain
\begin{align}\label{eq:MSE_KF_matteo}
\text{MSE}(\mathbf{f}_t^{fKF})=\mathbf P_t=&\,\left(\mathbf{\Phi P}_{t-1} \mathbf{\Phi}^{\prime}+ \mathbf{\Sigma}_{\eta}\right)\\
&-
 \left(\mathbf{\Lambda}^{\prime} \mathbf{\Sigma}_{\varepsilon}^{-1} \mathbf{\Lambda}+\left(\mathbf{\Phi P}_{t-1} \mathbf{\Phi}^{\prime}+ \mathbf{\Sigma}_{\eta}\right)^{-1}\right)^{-1}\mathbf{\Lambda}^{\prime} \mathbf{\Sigma}_{\varepsilon}^{-1} \mathbf{\Lambda}  \left(\mathbf{\Phi P}_{t-1} \mathbf{\Phi}^{\prime} + \mathbf{\Sigma}_{\eta}\right).\nonumber
\end{align}
Note that, unless $\mathbf{\Phi}$ is diagonal, the MSE matrix in (\ref{eq:MSE_KF_matteo}) is not diagonal. Moreover, the MSE of the Kalman filter estimates of the factors does not have a close form solution.



Several properties of the filtered estimates of the factors in (\ref{eq:KF_matteo}) are interesting for the purposes of this paper. First, given the assumptions on the factors and loadings, the state space representation of the DFM is stabilizable and detectable; see Lemma I.1 of Barigozzi and Luciani (2024). Consequently, the KF reaches its steady state exponentially fast, with the one-step-ahead and filtered MSEs  and, thus, also the Kalman gain in \eqref{eq:KF_matteo} becoming constant over time, i.e. $\lim_{t\to\infty}\mathbf{P}_{t|t-1}= \bar{\mathbf{\Pi}}$, $\lim_{t\to\infty}\mathbf{P}_t= \bar{\mathbf{P}}$, and $\lim_{t\to\infty}\mathbf{K}_t= \bar{\mathbf{K}}$ (see Anderson and Moore, 2005).
In this case, the MSE in \eqref{eq:MSE_KF_matteo} converges to 
\begin{align}\label{eq:MSE_KF_matteo_ss}
\lim_{t\to\infty} \text{MSE}(\mathbf{f}_t^{fKF})= \bar{\mathbf{ P}}=&\,\left(\mathbf{\Phi }  \bar{\mathbf{ P}} \mathbf{\Phi}^{\prime}+ \mathbf{\Sigma}_{\eta}\right)\\
&- \left(\mathbf{\Lambda}^{\prime} \mathbf{\Sigma}_{\varepsilon}^{-1} \mathbf{\Lambda}+ \left(\mathbf{\Phi} \bar{\mathbf P} \mathbf{\Phi}^{\prime}+ \mathbf{\Sigma}_{\eta}\right)^{-1}\right)^{-1}\mathbf{\Lambda}^{\prime} \mathbf{\Sigma}_{\varepsilon}^{-1} \mathbf{\Lambda}  \left(\mathbf{\Phi} \bar{\mathbf{ P}} \mathbf{\Phi}^{\prime} + \mathbf{\Sigma}_{\eta}\right),\nonumber
\end{align}
which, although it still does not have a closed form expression in general, it is no more time dependent.\footnote{Explicit expressions for the steady state MSEs and Kalman gain can be derived by solving a Riccati equation implied by the Kalman filter recursions. There is a large literature on algebraic as well as iterative solutions of the Riccati equation; see the proposals by Assimakis and Adam (2014) and the references therein. Poncela and Ruiz (2015) solve the Riccati equation for $r=1$. Note that if the autoregressive matrix $\Phi$ is assumed to be diagonal, then the solution of the Riccati equation for $r=1$ is the solution for each of the factors. Solving the Riccati equation for general $r$ and $\Phi$ is, however, beyond our objectives and left for further research. } The exact number of periods needed to reach the steady-state depends on $\mathbf \Phi$ and the initial conditions of the filter. In general, since the convergence rate is exponential in $t$, this happens within few time periods. Therefore \eqref{eq:MSE_KF_matteo_ss} can be viewed as a good approximation of the MSE of the fKF estimator.


Second, note that, at each fixed point in time $t$, the factors $\mathbf{f}_t^{fKF}$ are a weighted average of past and contemporaneous observations with the weights of these two components depending on the factor loadings, $\mathbf{\Lambda}$, the autoregressive parameters, $\mathbf{\Phi}$, and the Kalman gain, $\mathbf{{K}}_t$, which also depends on the idiosyncratic covariance matrix, $\mathbf{\Sigma}_{\varepsilon}$. In fact, if the factors were serially uncorrelated, then $\mathbf{\Phi}=\mathbf{0}_r$ and $\mathbf P_{t|t-1}=\mathbf{\Sigma}_{\eta}=\mathbf{\Sigma}_{F}=\mathbf{I}_r$ because of our identifying assumptions. Consequently, $\mathbf{K}_t=\mathbf{\Lambda}^{\prime} \left( \mathbf{\Lambda \Lambda}^{\prime} + \mathbf{\Sigma}_{\varepsilon}\right)^{-1}$, i.e., it does not depend on time, and the estimated factors only depend on contemporaneous observations, $\mathbf{Y}_t$. As expected, in this case, the filtered factors coincide with the linear projection $\mathbf{f}_t^{fLP}$ in (\ref{eq:LP_1}) with MSE given as in (\ref{eq:MSE_LP_2}). 

Third, since, as implied by assumption 1(c), when $N\to\infty$, $\mathbf{\Lambda}^\prime\mathbf{\Sigma}_{\varepsilon}^{-1}\mathbf \Lambda$ diverges as $O(N)$, then according to \eqref{eq:gain_3}, 
$\mathbf K_t$ tends to $(\mathbf{\Lambda}^\prime\mathbf{\Sigma}_{\varepsilon}^{-1}\mathbf \Lambda)^{-1}\mathbf{\Lambda}^\prime\mathbf{\Sigma}_{\varepsilon}^{-1}$ with an error $O(N^{-1})$ and, therefore,
$\mathbf K_t\mathbf \Lambda$ tends to $\mathbf I_r$. So from (\ref{eq:KF_matteo}) we see that the MSE of $\mathbf{f}_t^{fKF}$, which is $\mathbf P_t$, tends to a matrix of zeros as $O(N^{-1})$. Moreover, we immediately see that the fKF and GLS estimators are asymptotically equivalent, as $N\to\infty$, with an error $O(N^{-1})$.
Therefore, $\mathbf{f}_t^{fKF}$ is $\sqrt N$--consistent, by Chebychev's inequality, and it has the same asymptotic distribution as the one derived for the GLS in \eqref{eq:Asym_GLS}. As argued for the fLP estimator, in this case, such asymptotic statements should be interpreted as conditional on a realization $\mathbf F_1,\ldots, \mathbf F_T$. In this case, using the information about the past evolution of $\mathbf{Y}_t$ does not affect the MSE of $\mathbf{f}_t^{fKF}$, at least asymptotically, as $N\to\infty$.

Last,
\begin{align}\label{eq:biasKF}
E[\mathbf f_t^{fKF}|\mathbf F_t] &= E[\mathbf f_{t|t-1}^{fKF}|\mathbf F_t] + \mathbf K_t \left(\mathbf \Lambda\mathbf F_t- \mathbf \Lambda E[\mathbf f_{t|t-1}^{fKF}|\mathbf F_t]\right) + \mathbf K_t E[\boldsymbol\varepsilon_t|\mathbf F_t]\nonumber\\
&= \mathbf F_t +O\left(\frac 1N\right)+ \left( \mathbf{\Lambda}^{\prime} \mathbf{\Sigma}^{-1}_{\varepsilon} \mathbf{\Lambda} \right) ^{-1} \mathbf{\Lambda}^{\prime} \mathbf{\Sigma}^{-1}_{\varepsilon}  E[\boldsymbol\varepsilon_t|\mathbf F_t],
\end{align}
so if $E[\boldsymbol{\varepsilon}_t|\mathbf F_t]=\mathbf 0$, then the fKF is asymptotically unbiased and the MSE in \eqref{eq:MSE_KF_matteo} multiplied by $N$ is still a consistent estimator of the asymptotic variance in \eqref{eq:Asym_GLS}. As mentioned above for the fLP estimator, sufficient conditions would be that either $\boldsymbol{\varepsilon}_t$ and $\mathbf F_t$ are independent or $\mathbf Y_t$ and $\mathbf F_t$ are jointly normal.\footnote{Alternative proofs of unbiasedness of the KF, holding for any fixed $N$, but treating factors a fixed parameters, can be found in Duncan and Horn (1972) and Humpherys \textit{et al.} (2012). }
 

As in previous sections, we also consider two potential mis-specifications of the idiosyncratic covariance matrix when running the KF. First, we consider extracting the factors when wrongly considering the idiosyncratic components as if they were cross-sectionally uncorrelated, i.e., when using  $\mathbf{\Sigma}_{\varepsilon}^{*}=\text{diag}(\mathbf{\Sigma}_{\varepsilon})$ in place of $\mathbf{\Sigma}_{\varepsilon}$ when implementing the Kalman filter. Then, the estimated factors, denoted as $\mathbf{f}_t^{dKF}$, are estimated using the recursions in (\ref{eq:KF_matteo}) when computing the Kalman gain using $\mathbf{\Sigma}_{\varepsilon}^{*}$ instead of $\mathbf{\Sigma}_{\varepsilon}$, with
%
$\mathbf{P}_{t|t-1}^{(0)} = \mathbf{\Phi P}_{t-1}^{(0)} \mathbf{\Phi}^{\prime}+ \mathbf{\Sigma}_{\eta}$,
and with the corresponding conditional MSE, denoted by $\mathbf{P}^{(0)}_t$, calculated as in (\ref{eq:MSE_KF_matteo}) with  $\mathbf{\Sigma}_{\varepsilon}^{*}$ instead of $\mathbf{\Sigma}_{\varepsilon}$. As before $\mathbf{P}^{(0)}_t$ does not depend on $\mathbf Y_1,\ldots, \mathbf Y_t$, hence, it is also an unconditional MSE. Specifically,
\begin{align}\label{eq:wrong_matteo}
\text{MSE}^{(0)}(\mathbf f_t^{dKF})=\mathbf{P}_t^{(0)} =&\, \left(\mathbf{\Phi}\mathbf{P}_{t-1}^{(0)} \mathbf{\Phi}^{\prime}+ \mathbf{\Sigma}_{\eta}\right)\\
&- \left (\mathbf \Lambda^\prime\mathbf \Sigma_{\varepsilon}^{*-1} \mathbf \Lambda+\mathbf P_{t|t-1}^{(0)-1}\right)^{-1}\mathbf \Lambda^\prime\mathbf \Sigma_{\varepsilon}^{*-1} \mathbf{\Lambda} 
\left(\mathbf{\Phi P}_{t-1}^{(0)} \mathbf{\Phi}^{\prime} + \mathbf{\Sigma}_{\eta}\right),\nonumber
\end{align}
which, however, would be the true MSE of $\mathbf{f}^{dKF}_t$ only if the idiosyncratic covariance matrix were truly diagonal.

If, instead, $\mathbf{\Sigma}_{\varepsilon}$ is full, the true MSE of $\mathbf{f}_t^{dKF}$ can be obtained using the results by Harvey and Delle Monache (2009) as follows
\begin{align}
\label{eq:MSE_wrongKF_2}
\text{MSE}(\mathbf{f}_t^{dKF})=&\, \mathbf{P}_t= \, \Big(\mathbf{\Phi} \mathbf{P}_{t-1} \mathbf{\Phi}^{\prime} + \mathbf{\Sigma}_{\eta}\Big)  \\
&+\left(\mathbf{\Phi} \mathbf{P}^{(0)}_{t-1} \mathbf{\Phi}^{\prime} + \mathbf{\Sigma}_{\eta}\right)  \mathbf{\Lambda}^{\prime} \left(\mathbf{\Lambda} \left(\mathbf{\Phi} \mathbf{P}^{(0)}_{t-1} \mathbf{\Phi}^{\prime} + \mathbf{\Sigma}_{\eta}\right) \mathbf{\Lambda}^{\prime} + \mathbf{\Sigma}_{\varepsilon}^* \right)^{-1}\nonumber\\
&\cdot \Big(\mathbf{\Lambda} \left(\mathbf{\Phi} \mathbf{P}_{t-1} \mathbf{\Phi}^{\prime} + \mathbf{\Sigma}_{\eta}\right) \mathbf{\Lambda}^{\prime} + \mathbf{\Sigma}_{\varepsilon} \Big) \nonumber\\
&\cdot\left(\mathbf{\Lambda} \left(\mathbf{\Phi} \mathbf{P}^{(0)}_{t-1} \mathbf{\Phi}^{\prime} + \mathbf{\Sigma}_{\eta}\right) \mathbf{\Lambda}^{\prime} + \mathbf{\Sigma}_{\varepsilon}^* \right)^{-1} \mathbf{\Lambda} \left(\mathbf{\Phi} \mathbf{P}^{(0)}_{t-1} \mathbf{\Phi}^{\prime} + \mathbf{\Sigma}_{\eta}\right)\nonumber\\
& - \left(\mathbf{\Phi} \mathbf{P}^{(0)}_{t-1} \mathbf{\Phi}^{\prime} + \mathbf{\Sigma}_{\eta}\right) \mathbf{\Lambda}^{\prime}  \left(\mathbf{\Lambda} \left(\mathbf{\Phi} \mathbf{P}^{(0)}_{t-1} \mathbf{\Phi}^{\prime} + \mathbf{\Sigma}_{\eta}\right) \mathbf{\Lambda}^{\prime} + \mathbf{\Sigma}_{\varepsilon}^* \right)^{-1} \mathbf{\Lambda}  
\Big(\mathbf{\Phi} \mathbf{P}_{t-1} \mathbf{\Phi}^{\prime} + \mathbf{\Sigma}_{\eta}\Big)\nonumber\\
& - 
\Big(\mathbf{\Phi} \mathbf{P}_{t-1} \mathbf{\Phi}^{\prime} + \mathbf{\Sigma}_{\eta}\Big)
\mathbf{\Lambda}^{\prime}  
\left(\mathbf{\Lambda} \left(\mathbf{\Phi} \mathbf{P}^{(0)}_{t-1} \mathbf{\Phi}^{\prime} + \mathbf{\Sigma}_{\eta}\right) \mathbf{\Lambda}^{\prime} + \mathbf{\Sigma}_{\varepsilon}^* \right)^{-1} 
\mathbf{\Lambda}  
\left(\mathbf{\Phi} \mathbf{P}^{(0)}_{t-1} \mathbf{\Phi}^{\prime} + \mathbf{\Sigma}_{\eta}\right) 
,\nonumber
\end{align}
which is not diagonal, it does not have a closed form solution, and it is time dependent.

Second, denoting by $\mathbf{f}_t^{sKF}$ the factors extracted by the Kalman filter when wrongly treating the idiosyncratic covariance matrix as if it were spherical, i.e., when $\mathbf{\Sigma}_{\varepsilon}$ is replaced by $\sigma^2_{\varepsilon}\mathbf I_N$. In this case, 
the true MSE is obtained as in \eqref{eq:MSE_wrongKF_2}, but when replacing $\mathbf{\Sigma}^*_\varepsilon$ with $\sigma^2_\varepsilon\mathbf I_N$, that is,
\begin{align}
\label{eq:MSE_wrongKF_3}
\text{MSE}(\mathbf{f}_t^{sKF})=&\,  \mathbf{P}_t = \, \Big(\mathbf{\Phi} \mathbf{P}_{t-1} \mathbf{\Phi}^{\prime} + \mathbf{\Sigma}_{\eta}\Big)  \\
&+\left(\mathbf{\Phi} \mathbf{P}^{(0)}_{t-1} \mathbf{\Phi}^{\prime} + \mathbf{\Sigma}_{\eta}\right)  \mathbf{\Lambda}^{\prime} \left(\mathbf{\Lambda} \left(\mathbf{\Phi} \mathbf{P}^{(0)}_{t-1} \mathbf{\Phi}^{\prime} + \mathbf{\Sigma}_{\eta}\right) \mathbf{\Lambda}^{\prime} +\sigma^2_\varepsilon\mathbf I_N \right)^{-1}\nonumber\\
&\cdot \Big(\mathbf{\Lambda} \left(\mathbf{\Phi} \mathbf{P}_{t-1} \mathbf{\Phi}^{\prime} + \mathbf{\Sigma}_{\eta}\right) \mathbf{\Lambda}^{\prime} + \mathbf{\Sigma}_{\varepsilon} \Big) \nonumber\\
&\cdot\left(\mathbf{\Lambda} \left(\mathbf{\Phi} \mathbf{P}^{(0)}_{t-1} \mathbf{\Phi}^{\prime} + \mathbf{\Sigma}_{\eta}\right) \mathbf{\Lambda}^{\prime} + \sigma^2_\varepsilon\mathbf I_N \right)^{-1} \mathbf{\Lambda} \left(\mathbf{\Phi} \mathbf{P}^{(0)}_{t-1} \mathbf{\Phi}^{\prime} + \mathbf{\Sigma}_{\eta}\right)\nonumber\\
& - \left(\mathbf{\Phi} \mathbf{P}^{(0)}_{t-1} \mathbf{\Phi}^{\prime} + \mathbf{\Sigma}_{\eta}\right) 
\mathbf{\Lambda}^{\prime}  \left(\mathbf{\Lambda} \left(\mathbf{\Phi} \mathbf{P}^{(0)}_{t-1} \mathbf{\Phi}^{\prime} + \mathbf{\Sigma}_{\eta}\right) \mathbf{\Lambda}^{\prime} + \sigma^2_\varepsilon\mathbf I_N \right)^{-1} \mathbf{\Lambda} 
 \Big(\mathbf{\Phi} \mathbf{P}_{t-1} \mathbf{\Phi}^{\prime} + \mathbf{\Sigma}_{\eta}\Big)\nonumber\\
 & - 
  \Big(\mathbf{\Phi} \mathbf{P}_{t-1} \mathbf{\Phi}^{\prime} + \mathbf{\Sigma}_{\eta}\Big)
\mathbf{\Lambda}^{\prime}  \left(\mathbf{\Lambda} \left(\mathbf{\Phi} \mathbf{P}^{(0)}_{t-1} \mathbf{\Phi}^{\prime} + \mathbf{\Sigma}_{\eta}\right) \mathbf{\Lambda}^{\prime} + \sigma^2_\varepsilon\mathbf I_N \right)^{-1} \mathbf{\Lambda} 
\left(\mathbf{\Phi} \mathbf{P}^{(0)}_{t-1} \mathbf{\Phi}^{\prime} + \mathbf{\Sigma}_{\eta}\right) 
,\nonumber
\end{align}
with $\mathbf{P}^{(0)}_t$ calculated as in (\ref{eq:wrong_matteo}) with  $\sigma^2_\varepsilon\mathbf I_N$ instead of $\mathbf{\Sigma}_{\varepsilon}^*$. Again the MSE in \eqref{eq:MSE_wrongKF_3} is not diagonal, it does not have a closed form solution, and it is time dependent. 

The same comments made for the MSE of the fKF apply to the MSE of the dKF and sKF  in \eqref{eq:MSE_wrongKF_2} and \eqref{eq:MSE_wrongKF_3}. Namely, at the steady state we can derive an expression which does not depend on time by replacing in \eqref{eq:MSE_wrongKF_2} and \eqref{eq:MSE_wrongKF_3}, $\mathbf{P}_{t-1}^{(0)}$ with its steady state $\bar {\mathbf{P}}^{(0)}=\lim_{t\to\infty}\mathbf{P}_{t-1}^{(0)}$, where convergence is exponentially fast in $t$. Second, in the case of serially uncorrelated factors, i.e., when $\mathbf P_{t|t-1}^{(0)}=\mathbf{\Sigma}_{\eta}=\mathbf{\Sigma}_{F}=\mathbf{I}_r$ because of our identifying assumptions,  the dKF and sKF estimators coincide with the linear projections $\mathbf{f}_t^{dLP}$ and $\mathbf{f}_t^{sLP}$ in (\ref{eq:LPD}) and (\ref{eq:SLP}) and with MSEs given as in (\ref{eq:MSE_DLP_2}) and (\ref{eq:MSE_SLP_2}). 

Third, using the same arguments used for the dLP and sLP estimators in Appendix A, we can easily prove that the MSEs of $\mathbf{f}_t^{dKF}$ and $\mathbf{f}_t^{sKF}$ tend to a matrix of zeros as $O(N^{-1})$ and the dKF (sKF) and WLS (OLS) estimators are asymptotically equivalent, as $N\to\infty$, again with an error $O(N^{-1})$. Therefore, both the $\mathbf{f}_t^{fKF}$ and $\mathbf{f}_t^{sKF}$ are $\sqrt N$--consistent, by Chebychev's inequality, and they have the same asymptotic distribution as the one derived for the WLS in \eqref{eq:Asym_WLS} and for the OLS in \eqref{eq:Asym_LS}, respectively. Once more such asymptotic statements should be interpreted as conditional on a realization $\mathbf F_1,\ldots, \mathbf F_T$. 

Last, following the similar arguments as in \eqref{eq:biasKF}, it is straightforward to see that, if $E[\boldsymbol{\varepsilon}_t|\mathbf F_t]=\mathbf 0$, then the dKF and sKF are asymptotically unbiased and then their MSEs in \eqref{eq:MSE_wrongKF_2} and \eqref{eq:MSE_wrongKF_3} multiplied by $N$ are consistent estimators of the asymptotic variances in \eqref{eq:Asym_WLS} and \eqref{eq:Asym_LS}. 

\section{The effect of unknown parameters}\label{sec:param}

The considered estimators require in practice to estimate also the loadings, the idiosyncratic variances and covariances, as well as, for the KF methods, also the VAR parameters governing the dynamics of the factors. Here we review the main results in the literature showing that if $T$ is large enough then the estimation error of the parameters, which depends on $T$, is negligible with respect to the estimation error of the factors in the case of known parameters, which depends on $N$.

First, as already noticed, the correctly specified estimators, WLS, fLP, and fKF are almost never computed in practice  as they all require a regularized and positive definite estimator of the full idiosycnratic covariance matrix. The only exception is Bai and Liao (2016) where, for given QML estimators of $\mathbf \Lambda$ and $\mathbf {\Sigma}_\varepsilon$  (regularized via $L1$ penalty), they consider the GLS estimator of the factors. However, they do not provide any rates of convergence of such an estimator, although it is clear that such rates would depend not only on $N$ and $T$ but also on the degree of regularization.
Hence, it is not straightforward to derive the impact of the estimated parameters on the WLS, fLP, and fKF. 

Second, since the OLS estimator depends only on the loadings, it is the most commonly used, although not necessarily the most efficient, estimator. The common practice is to estimate $\mathbf{\Lambda}$ via PC. Then, the results in Bai (2003) imply that the PC estimated factors are such that $\mathbf f_t^{PC}=\mathbf f_t^{OLS}+O_p(T^{-1})$.\footnote{Bai (2003) writes the estimated factors as $\sqrt T$ times eigenvectors of a $T\times T$ covariance matrix but these are numerically identical to the linear projection of the data $\mathbf Y_t$ onto the  loadings estimated as eigenvectors of the $N\times N$ sample covariance matrix $T^{-1}\sum_{t=1}^T \mathbf Y_t\mathbf Y_t^{\prime}$ times the square-root  of the corresponding eigenvalues (see, e.g., Bai and Ng, 2020). Hence, all results in Bai (2003) can be derived equivalently by exchanging loadings and factors as well as the cross-sectional and time dimensions. Therefore, since Bai (2003) proves that the estimated loadings satisfy $\boldsymbol {\lambda}_i^{PC}=\boldsymbol{\lambda}_i^{OLS}+O(N^{-1})$, the stated result of the factors  holds too. } Therefore, all our derivations for the OLS hold also when estimating the loadings via PC provided that $\sqrt N/T\to 0$, as $N,T\to\infty$. We conjecture that the same would hold if the loadings were estimated via QML.

Third, the mis-specified estimators  WLS and dLP require estimation of the loadings and the idiosyncratic variances, which are typically estimated via QML. Let us denote the resulting estimators of the factors as $\mathbf f_t^{WLS/QML}$ and $\mathbf f_t^{dLP/QML}$. Then, Bai and Li (2016) show that 
$\mathbf f_t^{WLS/QML}=\mathbf f_t^{WLS}+O_p(T^{-1})$. Moreover, they also show that $\mathbf f_t^{dLP/QML}=\mathbf f_t^{WLS/QML}+O_p(N^{-1})$. 
If the loadings and idiosyncratic variances were estimated via PC, then we would have the WPC estimator studied by Breitung and Tenhofen (2011), who also show that
$\mathbf f_t^{WPC}=\mathbf f_t^{WLS}+O_p(T^{-1})$. Therefore, all our derivations for the WLS hold also when estimating the loadings and the idiosyncratic variances via QML or PC, provided that $\sqrt N/T\to 0$, as $N,T\to\infty$. As far as the dLP is concerned the same condition is sufficient if we estimate the loadings and the idiosyncratic variances via QML and we conjecture it will hold also if these were estimated via PC. Similar comments would hold for the mis-specified estimator sLP which, however, is never found in practice as it is probably too simple with respect to its more realistic counterpart dLP. 

Fourth, the dKF requires estimating the loadings, the idiosyncratic variances, and the VAR parameters, $\mathbf \Phi$ and $\mathbf \Sigma_\eta$. The typical approach here is to estimate the parameters via QML implemented via the Expectation Maximization (EM) algorithm which in turn requires to estimate the factors via Kalman smoother which, in turn, requires also computing the Kalman filter (see Doz \textit{et al.}, 2012).  Let us denote the resulting estimator of the factors via the Kalman filter and smoother as $\mathbf f_t^{dKF/EM}$  and $\mathbf f_t^{dKS/EM}$, respectively. Then, Barigozzi and Luciani (2024) show that $\mathbf f_t^{dKS/EM}=\mathbf f_t^{dKF/EM}+O_p(N^{-1})$ and $\mathbf f_t^{dKF/EM}=\mathbf f_t^{dKF}+O_p(T^{-1})$. Therefore, all our derivations for the dKF hold also when estimating the loadings, the idiosyncratic variances, and the VAR parameters via EM, provided that $\sqrt N/T\to 0$, as $N,T\to\infty$. We conjecture that the same would hold if the loadings and idiosyncratic variances were estimated via PC and the VAR parameters were estimated via OLS on the PC estimator of the factors. Similar comments would hold for the mis-specified estimator sKF which, however, is never found in practice as it is probably too simple with respect to its more realistic counterpart dLP.

\section{Numerical illustration}
\label{sec:Simu}

In this section, we illustrate the results above by comparing the theoretical finite sample MSEs derived in this paper for the different estimators considered and under different specifications of the DFM in (\ref{eq:DFM}). We also carry out Monte Carlo experiments to analyse whether those theoretical finite sample MSEs are a good approximation to the empirical MSEs. Finally, we illustrate the usefulness of the derived MSEs by constructing confidence intervals for a simulated factor using the different approaches described in this paper.

We consider the  DFM in (\ref{eq:DFM}) with one factor, $r=1$, and cross-sectional dimension $N\in\{50, 150, 500\}$.\footnote{Note hat the temporal size is not relevant due to the model parameters being known.} To obtain the parameters of the DFM, the loadings are generated by $\lambda_{1i} \sim U(0, 1)$, for $i=1,\ldots,N$, while we consider different specifications of the factor and of the idiosyncratic components. In particular, the following AR(1) process is assumed for the factor
\begin{equation}
\label{eq:factorAR}
F_t=\phi F_{t-1}+\eta_t,
\end{equation}
where $\eta_t$ is a Gaussian white noise with variance $1-\phi^2$. Therefore, the factor has zero mean and variance one.\footnote{We do need to impose $\mathbf \Lambda^\prime \mathbf{\Sigma}_{\varepsilon}^{-1} \mathbf\Lambda$ to be diagonal, since with just one factor this is a scalar quantity.} We consider $\phi\in\{0, 0.7, 0.97\}$ to take into account different degrees of serial dependence and persistence. The idiosyncratic components, are always simulated as Gaussian white noises, with the elements in the main diagonal of their covariance matrix, $\mathbf{\Sigma}_{\varepsilon}$,  given by $\sigma_i^2=\sigma^{*2}v_i$, while the off-diagonal elements, $\sigma_{ij}$, are generated by permuting the columns obtained according to the following Toeplitz structure $\sigma_{ij}=\sigma_i \sigma_j \tau^{|i-j|},  i=1,\ldots,N, \text{and } j=i+1,\ldots,N$. Hence, $v_i$, which is either 1 or random draws from an $U \sim (0.5,10)$, controls heteroscedasticity and $\tau \in \{0, 0.5\}$ controls correlation.
Last we consider $\sigma^{2*}\in\{0.5, 1,2\}$  to control for the ``noise-to-signal'' ratio. 

\subsection{Theoretical and empirical MSEs}
In Tables \ref{Tab2}-\ref{Tab5} we report the finite sample theoretical  MSEs (rows denoted with A) for the nine factor extraction  procedures described above, namely, ${f}_t^{GLS}$, ${f}_t^{WLS}$, ${f}_t^{OLS}$, ${f}_t^{fLP}$,  ${f}_t^{dLP}$,  ${f}_t^{sLP}$, ${f}_t^{fKF}$,  ${f}_t^{dKF}$, and ${f}_t^{sKF}$. In particular, we consider the following four specifications:
\begin{enumerate}
\item the factor is white noise and the idiosyncratic components are cross-sectionally homoscedastic and uncorrelated, i.e., $\phi=0$, $v_i=1$, and $\tau=0$;
\item the factor is white noise and the  idiosyncratic components are heteroscedastic and cross-sectionally uncorrelated, i.e., $\phi=0$, $v_i\sim U(0.5, 10)$, and $\tau=0$;
\item the factor is white noise and the idiosyncratic components are heteroscedastic and cross-sectionally correlated, i.e., $\phi=0$, $v_i\sim U(0.5, 10)$, and $\tau=0.5$;
\item the factor is autocorrelated and the idiosyncratic components are heteroscedastic and cross-sectionally correlated, i.e., $\phi=0.7$, $v_i\sim U(0.5, 10)$, and $\tau=0.5$.
\end{enumerate}

Several interesting conclusions can be obtained from Tables \ref{Tab2}-\ref{Tab5}. First, as expected, in all designs in which the factors are white noise, the factors extracted using LP and KF coincide and, consequently, they have the same finite sample MSEs. Furthermore, in this case, if the idiosyncratic components are cross-sectionally homoscedastic and uncorrelated, the finite sample MSEs of the factors extracted by all procedures coincide regardless of $N$ when $\sigma^{2*}$ is small. However, as $\sigma^{2*}$ increases, the procedures based on LP are more efficient than those based on LS for small $N$. If $N$ is large, the finite sample MSEs of LS and filtered estimates of the factors are the same.

Cross-sectional heteroscedasticity of the idiosyncratic components has two main impacts on the finite sample MSEs of the factors. First, there is a large loss of efficiency of the procedures that assume constant variance, which have MSEs that nearly double those of the corresponding procedures that assume diagonal $\mathbf{\Sigma}_{\varepsilon}$. Furthermore, even if $\sigma^{2*}$ is small, the procedures based on filtering are more efficient than those based on LS. If the idiosyncratic components are further cross-sectionally correlated, the difference between the finite sample MSEs of the procedures that assume diagonal $\mathbf{\Sigma}_{\varepsilon}$ and those that assume it is full, can be substantial when the cross-sectional dimension, $N$, is large and the idiosyncratic variances are sizeable.

Summarising, if the factors are white noise, for large $N$, factor extraction procedures based on LS are as efficient as those based on filtering. However, for small and moderate $N$, filtering is more efficient than LS with the difference between both procedures being larger as the variances of the idiosyncratic components become substantial and/or their covariance matrix is not scalar.

Finally, note that the finite sample MSEs of the LS and LP factors are exactly the same regardless of whether the factor has temporal dependence or it is white noise. However, by using the KF in this case, we reduce the MSE with this reduction being larger the larger the variance of the idiosyncratic components, and the smaller the cross-sectional dimension, $N$. 

\begin{table}[htbp]
\centering
\footnotesize{
\begin{tabular}{cccccccccccc}
\hline
\hline
$\sigma^{2\ast}$ & $N$ &  & \textbf{OLS} & \textbf{WLS} & \textbf{GLS} & \textbf{sLP} & \textbf{dLP} & \textbf{fLP} & \textbf{sKF} & \textbf{dKF} & \textbf{fKF}\tabularnewline
\hline
 & {\footnotesize\textbf{50}} & {\footnotesize A} & {\footnotesize 0.032} & {\footnotesize 0.032} & {\footnotesize 0.032} & {\footnotesize 0.031} & {\footnotesize 0.031} & {\footnotesize 0.031} & {\footnotesize 0.031} & {\footnotesize 0.031} & {\footnotesize 0.031}\\
\vspace{0.001cm}
 &  & {\footnotesize E} & {\footnotesize 0.032} & {\footnotesize 0.032} & {\footnotesize 0.032} & {\footnotesize 0.031} & {\footnotesize 0.031} & {\footnotesize 0.031} & {\footnotesize 0.031} & {\footnotesize 0.031} & {\footnotesize 0.031}\\
{\footnotesize\textbf{0.5}} & {\footnotesize\textbf{150}} & {\footnotesize A} & {\footnotesize 0.011} & {\footnotesize 0.011} & {\footnotesize 0.011} & {\footnotesize 0.011} & {\footnotesize 0.011} & {\footnotesize 0.011} & {\footnotesize 0.011} & {\footnotesize 0.011} & {\footnotesize 0.011}\tabularnewline
 &  & {\footnotesize E} & {\footnotesize 0.011} & {\footnotesize 0.011} & {\footnotesize 0.011} & {\footnotesize 0.011} & {\footnotesize 0.011} & {\footnotesize 0.011} & {\footnotesize 0.011} & {\footnotesize 0.011} & {\footnotesize 0.011}\tabularnewline
 & {\footnotesize\textbf{500}} & {\footnotesize A} & {\footnotesize 0.003} & {\footnotesize 0.003} & {\footnotesize 0.003} & {\footnotesize 0.003} & {\footnotesize 0.003} & {\footnotesize 0.003} & {\footnotesize 0.003} & {\footnotesize 0.003} & {\footnotesize 0.003}\tabularnewline
 &  & {\footnotesize E} & {\footnotesize 0.003} & {\footnotesize 0.003} & {\footnotesize 0.003} & {\footnotesize 0.003} & {\footnotesize 0.003} & {\footnotesize 0.003} & {\footnotesize 0.003} & {\footnotesize 0.003} & {\footnotesize 0.003}\tabularnewline
\hline 
 & {\footnotesize\textbf{50}} & {\footnotesize A} & {\footnotesize 0.064} & {\footnotesize 0.064} & {\footnotesize 0.064} & {\footnotesize 0.061} & {\footnotesize 0.061} & {\footnotesize 0.061} & {\footnotesize 0.061} & {\footnotesize 0.061} & {\footnotesize 0.061}\tabularnewline
 &  & {\footnotesize E} & {\footnotesize 0.064} & {\footnotesize 0.064} & {\footnotesize 0.064} & {\footnotesize 0.061} & {\footnotesize 0.061} & {\footnotesize 0.061} & {\footnotesize 0.061} & {\footnotesize 0.061} & {\footnotesize 0.061}\tabularnewline
{\footnotesize\textbf{1}} & {\footnotesize\textbf{150}} & {\footnotesize A} & {\footnotesize 0.022} & {\footnotesize 0.022} & {\footnotesize 0.022} & {\footnotesize 0.022} & {\footnotesize 0.022} & {\footnotesize 0.022} & {\footnotesize 0.022} & {\footnotesize 0.022} & {\footnotesize 0.022}\tabularnewline
 &  & {\footnotesize E} & {\footnotesize 0.022} & {\footnotesize 0.022} & {\footnotesize 0.022} & {\footnotesize 0.022} & {\footnotesize 0.022} & {\footnotesize 0.022} & {\footnotesize 0.022} & {\footnotesize 0.022} & {\footnotesize 0.022}\tabularnewline
 & {\footnotesize\textbf{500}} & {\footnotesize A} & {\footnotesize 0.006} & {\footnotesize 0.006} & {\footnotesize 0.006} & {\footnotesize 0.006} & {\footnotesize 0.006} & {\footnotesize 0.006} & {\footnotesize 0.006} & {\footnotesize 0.006} & {\footnotesize 0.006}\tabularnewline
 &  & {\footnotesize E} & {\footnotesize 0.006} & {\footnotesize 0.006} & {\footnotesize 0.006} & {\footnotesize 0.006} & {\footnotesize 0.006} & {\footnotesize 0.006} & {\footnotesize 0.006} & {\footnotesize 0.006} & {\footnotesize 0.006}\tabularnewline
\hline 
 & {\footnotesize\textbf{50}} & {\footnotesize A} & {\footnotesize 0.128} & {\footnotesize 0.128} & {\footnotesize 0.128} & {\footnotesize 0.114} & {\footnotesize 0.114} & {\footnotesize 0.114} & {\footnotesize 0.114} & {\footnotesize 0.114} & {\footnotesize 0.114}\tabularnewline
 &  & {\footnotesize E} & {\footnotesize 0.129} & {\footnotesize 0.129} & {\footnotesize 0.129} & {\footnotesize 0.114} & {\footnotesize 0.114} & {\footnotesize 0.114} & {\footnotesize 0.114} & {\footnotesize 0.114} & {\footnotesize 0.114}\tabularnewline
{\footnotesize\textbf{2}} & {\footnotesize\textbf{150}} & {\footnotesize A} & {\footnotesize 0.044} & {\footnotesize 0.044} & {\footnotesize 0.044} & {\footnotesize 0.042} & {\footnotesize 0.042} & {\footnotesize 0.042} & {\footnotesize 0.042} & {\footnotesize 0.042} & {\footnotesize 0.042}\tabularnewline
 &  & {\footnotesize E} & {\footnotesize 0.044} & {\footnotesize 0.044} & {\footnotesize 0.044} & {\footnotesize 0.042} & {\footnotesize 0.042} & {\footnotesize 0.042} & {\footnotesize 0.042} & {\footnotesize 0.042} & {\footnotesize 0.042}\tabularnewline
 & {\footnotesize\textbf{500}} & {\footnotesize A} & {\footnotesize 0.012} & {\footnotesize 0.012} & {\footnotesize 0.012} & {\footnotesize 0.012} & {\footnotesize 0.012} & {\footnotesize 0.012} & {\footnotesize 0.012} & {\footnotesize 0.012} & {\footnotesize 0.012}\tabularnewline
 &  & {\footnotesize E} & {\footnotesize 0.012} & {\footnotesize 0.012} & {\footnotesize 0.012} & {\footnotesize 0.012} & {\footnotesize 0.012} & {\footnotesize 0.012} & {\footnotesize 0.012} & {\footnotesize 0.012} & {\footnotesize 0.012}\tabularnewline
 \hline
 \hline
\end{tabular}
}

\caption{White noise factor, homoscedastic and uncorrelated idiosyncratic components, i.e, $\phi=0$, $v_i=1$, and $\tau=0$. Finite sample theoretical MSEs (A) and empirical MSEs (E).}
\label{Tab2}
\end{table}

\begin{table}[htbp]
\centering
\footnotesize{
\begin{tabular}{cccccccccccc}
\hline
\hline
$\sigma^{2\ast}$ & $N$ &  & \textbf{OLS} & \textbf{WLS} & \textbf{GLS} & \textbf{sLP} & \textbf{dLP} & \textbf{fLP} & \textbf{sKF} & \textbf{dKF} & \textbf{fKF}\tabularnewline
\hline
 & {\footnotesize\textbf{50}} & {\footnotesize A} & {\footnotesize 0.178} & {\footnotesize 0.094} & {\footnotesize 0.094} & {\footnotesize 0.151} & {\footnotesize 0.086} & {\footnotesize 0.086} & {\footnotesize 0.151} & {\footnotesize 0.086} & {\footnotesize 0.086}\tabularnewline
 &  & {\footnotesize E} & {\footnotesize 0.179} & {\footnotesize 0.094} & {\footnotesize 0.094} & {\footnotesize 0.152} & {\footnotesize 0.086} & {\footnotesize 0.086} & {\footnotesize 0.152} & {\footnotesize 0.086} & {\footnotesize 0.086}\tabularnewline
{\footnotesize\textbf{0.5}} & {\footnotesize\textbf{150}} & {\footnotesize A} & {\footnotesize 0.058} & {\footnotesize 0.036} & {\footnotesize 0.036} & {\footnotesize 0.055} & {\footnotesize 0.035} & {\footnotesize 0.035} & {\footnotesize 0.055} & {\footnotesize 0.035} & {\footnotesize 0.035}\tabularnewline
 &  & {\footnotesize E} & {\footnotesize 0.058} & {\footnotesize 0.036} & {\footnotesize 0.036} & {\footnotesize 0.055} & {\footnotesize 0.035} & {\footnotesize 0.035} & {\footnotesize 0.055} & {\footnotesize 0.035} & {\footnotesize 0.035}\tabularnewline
 & {\footnotesize\textbf{500}} & {\footnotesize A} & {\footnotesize 0.015} & {\footnotesize 0.009} & {\footnotesize 0.009} & {\footnotesize 0.015} & {\footnotesize 0.009} & {\footnotesize 0.009} & {\footnotesize 0.015} & {\footnotesize 0.009} & {\footnotesize 0.009}\tabularnewline
 &  & {\footnotesize E} & {\footnotesize 0.015} & {\footnotesize 0.009} & {\footnotesize 0.009} & {\footnotesize 0.015} & {\footnotesize 0.009} & {\footnotesize 0.009} & {\footnotesize 0.015} & {\footnotesize 0.009} & {\footnotesize 0.009}\tabularnewline
\hline 
 & {\footnotesize\textbf{50}} & {\footnotesize A} & {\footnotesize 0.356} & {\footnotesize 0.188} & {\footnotesize 0.188} & {\footnotesize 0.263} & {\footnotesize 0.158} & {\footnotesize 0.158} & {\footnotesize 0.263} & {\footnotesize 0.158} & {\footnotesize 0.158}\tabularnewline
 &  & {\footnotesize E} & {\footnotesize 0.358} & {\footnotesize 0.189} & {\footnotesize 0.189} & {\footnotesize 0.263} & {\footnotesize 0.159} & {\footnotesize 0.159} & {\footnotesize 0.263} & {\footnotesize 0.158} & {\footnotesize 0.158}\tabularnewline
{\footnotesize\textbf{1}} & {\footnotesize\textbf{150}} & {\footnotesize A} & {\footnotesize 0.116} & {\footnotesize 0.073} & {\footnotesize 0.073} & {\footnotesize 0.104} & {\footnotesize 0.068} & {\footnotesize 0.068} & {\footnotesize 0.104} & {\footnotesize 0.068} & {\footnotesize 0.068}\tabularnewline
 &  & {\footnotesize E} & {\footnotesize 0.116} & {\footnotesize 0.073} & {\footnotesize 0.073} & {\footnotesize 0.104} & {\footnotesize 0.068} & {\footnotesize 0.068} & {\footnotesize 0.104} & {\footnotesize 0.068} & {\footnotesize 0.068}\tabularnewline
 & {\footnotesize\textbf{500}} & {\footnotesize A} & {\footnotesize 0.030} & {\footnotesize 0.017} & {\footnotesize 0.017} & {\footnotesize 0.029} & {\footnotesize 0.017} & {\footnotesize 0.017} & {\footnotesize 0.029} & {\footnotesize 0.017} & {\footnotesize 0.017}\tabularnewline
 &  & {\footnotesize E} & {\footnotesize 0.030} & {\footnotesize 0.017} & {\footnotesize 0.017} & {\footnotesize 0.029} & {\footnotesize 0.017} & {\footnotesize 0.017} & {\footnotesize 0.029} & {\footnotesize 0.017} & {\footnotesize 0.017}\tabularnewline
\hline 
 & {\footnotesize\textbf{50}} & {\footnotesize A} & {\footnotesize 0.713} & {\footnotesize 0.376} & {\footnotesize 0.376} & {\footnotesize 0.416} & {\footnotesize 0.273} & {\footnotesize 0.273} & {\footnotesize 0.416} & {\footnotesize 0.273} & {\footnotesize 0.273}\tabularnewline
 &  & {\footnotesize E} & {\footnotesize 0.715} & {\footnotesize 0.377} & {\footnotesize 0.377} & {\footnotesize 0.417} & {\footnotesize 0.274} & {\footnotesize 0.274} & {\footnotesize 0.416} & {\footnotesize 0.273} & {\footnotesize 0.273}\tabularnewline
{\footnotesize\textbf{2}} & {\footnotesize\textbf{150}} & {\footnotesize A} & {\footnotesize 0.232} & {\footnotesize 0.145} & {\footnotesize 0.145} & {\footnotesize 0.188} & {\footnotesize 0.127} & {\footnotesize 0.127} & {\footnotesize 0.188} & {\footnotesize 0.127} & {\footnotesize 0.127}\tabularnewline
 &  & {\footnotesize E} & {\footnotesize 0.232} & {\footnotesize 0.145} & {\footnotesize 0.145} & {\footnotesize 0.188} & {\footnotesize 0.127} & {\footnotesize 0.127} & {\footnotesize 0.188} & {\footnotesize 0.127} & {\footnotesize 0.127}\tabularnewline
 & {\footnotesize\textbf{500}} & {\footnotesize A} & {\footnotesize 0.060} & {\footnotesize 0.035} & {\footnotesize 0.035} & {\footnotesize 0.057} & {\footnotesize 0.034} & {\footnotesize 0.034} & {\footnotesize 0.057} & {\footnotesize 0.034} & {\footnotesize 0.034}\tabularnewline
 &  & {\footnotesize E} & {\footnotesize 0.060} & {\footnotesize 0.035} & {\footnotesize 0.035} & {\footnotesize 0.056} & {\footnotesize 0.034} & {\footnotesize 0.034} & {\footnotesize 0.056} & {\footnotesize 0.034} & {\footnotesize 0.034}\tabularnewline
 \hline
 \hline
\end{tabular}
}

\caption{White noise factor, heteroscedastic and uncorrelated idiosyncratic components, i.e, $\phi=0$, $v_i\sim U(0.5, 10)$, and $\tau=0$. Finite sample theoretical MSEs (A) and empirical MSEs (E).}
\label{Tab3}
\end{table}

\begin{table}[htbp]
\centering
\footnotesize{
\begin{tabular}{cccccccccccc}
\hline
\hline
$\sigma^{2\ast}$ & $N$ &  & \textbf{OLS} & \textbf{WLS} & \textbf{GLS} & \textbf{sLP} & \textbf{dLP} & \textbf{fLP} & \textbf{sKF} & \textbf{dKF} & \textbf{fKF}\tabularnewline
\hline
 & {\footnotesize\textbf{50}} & {\footnotesize A} & {\footnotesize 0.191} & {\footnotesize 0.104} & {\footnotesize 0.101} & {\footnotesize 0.161} & {\footnotesize 0.095} & {\footnotesize 0.092} & {\footnotesize 0.161} & {\footnotesize 0.095} & {\footnotesize 0.092}\tabularnewline
 &  & {\footnotesize E} & {\footnotesize 0.192} & {\footnotesize 0.105} & {\footnotesize 0.101} & {\footnotesize 0.161} & {\footnotesize 0.095} & {\footnotesize 0.092} & {\footnotesize 0.161} & {\footnotesize 0.095} & {\footnotesize 0.092}\tabularnewline
{\footnotesize\textbf{0.5}} & {\footnotesize\textbf{150}} & {\footnotesize A} & {\footnotesize 0.079} & {\footnotesize 0.050} & {\footnotesize 0.041} & {\footnotesize 0.074} & {\footnotesize 0.047} & {\footnotesize 0.040} & {\footnotesize 0.074} & {\footnotesize 0.047} & {\footnotesize 0.040}\tabularnewline
 &  & {\footnotesize E} & {\footnotesize 0.079} & {\footnotesize 0.050} & {\footnotesize 0.041} & {\footnotesize 0.074} & {\footnotesize 0.047} & {\footnotesize 0.040} & {\footnotesize 0.074} & {\footnotesize 0.047} & {\footnotesize 0.040}\tabularnewline
 & {\footnotesize\textbf{500}} & {\footnotesize A} & {\footnotesize 0.035} & {\footnotesize 0.020} & {\footnotesize 0.011} & {\footnotesize 0.034} & {\footnotesize 0.020} & {\footnotesize 0.011} & {\footnotesize 0.034} & {\footnotesize 0.020} & {\footnotesize 0.011}\tabularnewline
 &  & {\footnotesize E} & {\footnotesize 0.035} & {\footnotesize 0.020} & {\footnotesize 0.011} & {\footnotesize 0.034} & {\footnotesize 0.020} & {\footnotesize 0.011} & {\footnotesize 0.034} & {\footnotesize 0.020} & {\footnotesize 0.011}\tabularnewline
\hline 
 & {\footnotesize\textbf{50}} & {\footnotesize A} & {\footnotesize 0.383} & {\footnotesize 0.209} & {\footnotesize 0.202} & {\footnotesize 0.278} & {\footnotesize 0.173} & {\footnotesize 0.168} & {\footnotesize 0.278} & {\footnotesize 0.173} & {\footnotesize 0.168}\tabularnewline
 &  & {\footnotesize E} & {\footnotesize 0.384} & {\footnotesize 0.209} & {\footnotesize 0.202} & {\footnotesize 0.278} & {\footnotesize 0.173} & {\footnotesize 0.168} & {\footnotesize 0.278} & {\footnotesize 0.173} & {\footnotesize 0.168}\tabularnewline
{\footnotesize\textbf{1}} & {\footnotesize\textbf{150}} & {\footnotesize A} & {\footnotesize 0.159} & {\footnotesize 0.099} & {\footnotesize 0.082} & {\footnotesize 0.138} & {\footnotesize 0.091} & {\footnotesize 0.076} & {\footnotesize 0.138} & {\footnotesize 0.091} & {\footnotesize 0.076}\tabularnewline
 &  & {\footnotesize E} & {\footnotesize 0.159} & {\footnotesize 0.099} & {\footnotesize 0.082} & {\footnotesize 0.138} & {\footnotesize 0.091} & {\footnotesize 0.076} & {\footnotesize 0.138} & {\footnotesize 0.091} & {\footnotesize 0.076}\tabularnewline
 & {\footnotesize\textbf{500}} & {\footnotesize A} & {\footnotesize 0.070} & {\footnotesize 0.040} & {\footnotesize 0.022} & {\footnotesize 0.067} & {\footnotesize 0.039} & {\footnotesize 0.021} & {\footnotesize 0.067} & {\footnotesize 0.039} & {\footnotesize 0.021}\tabularnewline
 &  & {\footnotesize E} & {\footnotesize 0.070} & {\footnotesize 0.040} & {\footnotesize 0.021} & {\footnotesize 0.067} & {\footnotesize 0.039} & {\footnotesize 0.021} & {\footnotesize 0.067} & {\footnotesize 0.039} & {\footnotesize 0.021}\tabularnewline
\hline 
 & {\footnotesize\textbf{50}} & {\footnotesize A} & {\footnotesize 0.766} & {\footnotesize 0.417} & {\footnotesize 0.404} & {\footnotesize 0.436} & {\footnotesize 0.295} & {\footnotesize 0.288} & {\footnotesize 0.436} & {\footnotesize 0.295} & {\footnotesize 0.288}\tabularnewline
 &  & {\footnotesize E} & {\footnotesize 0.769} & {\footnotesize 0.418} & {\footnotesize 0.405} & {\footnotesize 0.436} & {\footnotesize 0.295} & {\footnotesize 0.288} & {\footnotesize 0.435} & {\footnotesize 0.295} & {\footnotesize 0.287}\tabularnewline
{\footnotesize\textbf{2}} & {\footnotesize\textbf{150}} & {\footnotesize A} & {\footnotesize 0.317} & {\footnotesize 0.198} & {\footnotesize 0.165} & {\footnotesize 0.245} & {\footnotesize 0.167} & {\footnotesize 0.141} & {\footnotesize 0.245} & {\footnotesize 0.167} & {\footnotesize 0.141}\tabularnewline
 &  & {\footnotesize E} & {\footnotesize 0.318} & {\footnotesize 0.198} & {\footnotesize 0.165} & {\footnotesize 0.245} & {\footnotesize 0.167} & {\footnotesize 0.141} & {\footnotesize 0.245} & {\footnotesize 0.167} & {\footnotesize 0.141}\tabularnewline
 & {\footnotesize\textbf{500}} & {\footnotesize A} & {\footnotesize 0.140} & {\footnotesize 0.079} & {\footnotesize 0.043} & {\footnotesize 0.128} & {\footnotesize 0.075} & {\footnotesize 0.041} & {\footnotesize 0.128} & {\footnotesize 0.075} & {\footnotesize 0.041}\tabularnewline
 &  & {\footnotesize E} & {\footnotesize 0.141} & {\footnotesize 0.079} & {\footnotesize 0.043} & {\footnotesize 0.128} & {\footnotesize 0.075} & {\footnotesize 0.041} & {\footnotesize 0.128} & {\footnotesize 0.075} & {\footnotesize 0.041}\tabularnewline
 \hline
 \hline
\end{tabular}
}

\caption{White noise factor, heteroscedastic and correlated idiosyncratic components, i.e, $\phi=0$, $v_i\sim U(0.5, 10)$, and $\tau=0.5$. Finite sample theoretical MSEs (A) and empirical MSEs (E).}
\label{Tab4}
\end{table}

\begin{table}[htbp]
\centering
\footnotesize{
\begin{tabular}{cccccccccccc}
\hline
\hline
$\sigma^{2\ast}$ & $N$ &  & \textbf{OLS} & \textbf{WLS} & \textbf{GLS} & \textbf{sLP} & \textbf{dLP} & \textbf{fLP} & \textbf{sKF} & \textbf{dKF} & \textbf{fKF}\tabularnewline
\hline
 & {\footnotesize\textbf{50}} & {\footnotesize A} & {\footnotesize 0.191} & {\footnotesize 0.104} & {\footnotesize 0.101} & {\footnotesize 0.161} & {\footnotesize 0.095} & {\footnotesize 0.092} & {\footnotesize 0.136} & {\footnotesize 0.080} & {\footnotesize 0.085}\tabularnewline
 &  & {\footnotesize E} & {\footnotesize 0.192} & {\footnotesize 0.105} & {\footnotesize 0.101} & {\footnotesize 0.161} & {\footnotesize 0.095} & {\footnotesize 0.092} & {\footnotesize 0.146} & {\footnotesize 0.088} & {\footnotesize 0.086}\tabularnewline
{\footnotesize\textbf{0.5}} & {\footnotesize\textbf{150}} & {\footnotesize A} & {\footnotesize 0.079} & {\footnotesize 0.050} & {\footnotesize 0.041} & {\footnotesize 0.074} & {\footnotesize 0.047} & {\footnotesize 0.040} & {\footnotesize 0.066} & {\footnotesize 0.041} & {\footnotesize 0.038}\tabularnewline
 &  & {\footnotesize E} & {\footnotesize 0.079} & {\footnotesize 0.050} & {\footnotesize 0.041} & {\footnotesize 0.074} & {\footnotesize 0.047} & {\footnotesize 0.039} & {\footnotesize 0.070} & {\footnotesize 0.046} & {\footnotesize 0.038}\tabularnewline
 & {\footnotesize\textbf{500}} & {\footnotesize A} & {\footnotesize 0.035} & {\footnotesize 0.020} & {\footnotesize 0.011} & {\footnotesize 0.034} & {\footnotesize 0.020} & {\footnotesize 0.011} & {\footnotesize 0.032} & {\footnotesize 0.018} & {\footnotesize 0.011}\tabularnewline
 &  & {\footnotesize E} & {\footnotesize 0.035} & {\footnotesize 0.020} & {\footnotesize 0.011} & {\footnotesize 0.034} & {\footnotesize 0.020} & {\footnotesize 0.011} & {\footnotesize 0.034} & {\footnotesize 0.019} & {\footnotesize 0.011}\tabularnewline
\hline
 & {\footnotesize\textbf{50}} & {\footnotesize A} & {\footnotesize 0.383} & {\footnotesize 0.209} & {\footnotesize 0.202} & {\footnotesize 0.278} & {\footnotesize 0.173} & {\footnotesize 0.168} & {\footnotesize 0.226} & {\footnotesize 0.143} & {\footnotesize 0.150}\tabularnewline
 &  & {\footnotesize E} & {\footnotesize 0.384} & {\footnotesize 0.209} & {\footnotesize 0.202} & {\footnotesize 0.278} & {\footnotesize 0.173} & {\footnotesize 0.168} & {\footnotesize 0.242} & {\footnotesize 0.155} & {\footnotesize 0.151}\tabularnewline
{\footnotesize\textbf{1}} & {\footnotesize\textbf{150}} & {\footnotesize A} & {\footnotesize 0.159} & {\footnotesize 0.099} & {\footnotesize 0.082} & {\footnotesize 0.138} & {\footnotesize 0.091} & {\footnotesize 0.076} & {\footnotesize 0.119} & {\footnotesize 0.077} & {\footnotesize 0.071}\tabularnewline
 &  & {\footnotesize E} & {\footnotesize 0.159} & {\footnotesize 0.099} & {\footnotesize 0.082} & {\footnotesize 0.138} & {\footnotesize 0.091} & {\footnotesize 0.076} & {\footnotesize 0.127} & {\footnotesize 0.085} & {\footnotesize 0.072}\tabularnewline
 & {\footnotesize\textbf{500}} & {\footnotesize A} & {\footnotesize 0.070} & {\footnotesize 0.040} & {\footnotesize 0.022} & {\footnotesize 0.067} & {\footnotesize 0.039} & {\footnotesize 0.021} & {\footnotesize 0.062} & {\footnotesize 0.035} & {\footnotesize 0.021}\tabularnewline
 &  & {\footnotesize E} & {\footnotesize 0.070} & {\footnotesize 0.040} & {\footnotesize 0.021} & {\footnotesize 0.067} & {\footnotesize 0.039} & {\footnotesize 0.021} & {\footnotesize 0.065} & {\footnotesize 0.038} & {\footnotesize 0.021}\tabularnewline
\hline 
 & {\footnotesize\textbf{50}} & {\footnotesize A} & {\footnotesize 0.766} & {\footnotesize 0.417} & {\footnotesize 0.404} & {\footnotesize 0.436} & {\footnotesize 0.295} & {\footnotesize 0.288} & {\footnotesize 0.349} & {\footnotesize 0.242} & {\footnotesize 0.246}\tabularnewline
 &  & {\footnotesize E} & {\footnotesize 0.769} & {\footnotesize 0.418} & {\footnotesize 0.405} & {\footnotesize 0.436} & {\footnotesize 0.295} & {\footnotesize 0.287} & {\footnotesize 0.369} & {\footnotesize 0.255} & {\footnotesize 0.249}\tabularnewline
{\footnotesize\textbf{2}} & {\footnotesize\textbf{150}} & {\footnotesize A} & {\footnotesize 0.317} & {\footnotesize 0.198} & {\footnotesize 0.165} & {\footnotesize 0.245} & {\footnotesize 0.167} & {\footnotesize 0.141} & {\footnotesize 0.201} & {\footnotesize 0.137} & {\footnotesize 0.128}\tabularnewline
 &  & {\footnotesize E} & {\footnotesize 0.318} & {\footnotesize 0.198} & {\footnotesize 0.165} & {\footnotesize 0.245} & {\footnotesize 0.167} & {\footnotesize 0.141} & {\footnotesize 0.215} & {\footnotesize 0.151} & {\footnotesize 0.129}\tabularnewline
 & {\footnotesize\textbf{500}} & {\footnotesize A} & {\footnotesize 0.140} & {\footnotesize 0.079} & {\footnotesize 0.043} & {\footnotesize 0.128} & {\footnotesize 0.075} & {\footnotesize 0.041} & {\footnotesize 0.115} & {\footnotesize 0.066} & {\footnotesize 0.040}\tabularnewline
 &  & {\footnotesize E} & {\footnotesize 0.141} & {\footnotesize 0.079} & {\footnotesize 0.043} & {\footnotesize 0.128} & {\footnotesize 0.075} & {\footnotesize 0.041} & {\footnotesize 0.120} & {\footnotesize 0.072} & {\footnotesize 0.040}\tabularnewline
 \hline
 \hline
\end{tabular}
}

\caption{Autocorrelated factor, heteroscedastic and correlated idiosyncratic components, i.e, $\phi=0.7$, $v_i\sim U(0.5, 10)$, and $\tau=0.5$. Finite sample theoretical MSEs (A) and empirical MSEs (E).}
\label{Tab5}
\end{table}

We then carry out a Monte Carlo experiment to study the quality of the approximation of the MSEs reported in Tables \ref{Tab2}-\ref{Tab5} to the true empirical MSEs. For this goal, we generate $B=1000$ replications of the DFM in (\ref{eq:DFM}) with the same loadings (generated once for all replications) and factors described above, and idiosyncratic components, under the same four setting considered above. Tables \ref{Tab2}-\ref{Tab5} report the average through time of empirical MSEs (rows denoted with E), given by
\[
\frac 1B \sum_{b=1}^B (f_t^{(b)}-F_t^{(b)})^2,
\]
where $f_t^{(b)}$ is any of the the nine estimators considered in this paper, and $F_t^{(b)}$ is the true simulated factor, both at a given replication $b$. We can observe that, for all designs considered, the theoretical MSEs coincide with the empirical MSEs.

\subsection{Confidence intervals}
In a second exercise, we study the implications for the construction of 95\% confidence bands for the estimated factors based on their asymptotic distribution, when estimating the asymptotic variance with the corresponding MSE. i.e., given by
\[
f_t \pm 1.96 \sqrt{ \text{MSE}(f_t)},
\]
where $f_t$ is any of the the nine  estimators considered in this paper.\footnote{Here we have joint Gaussianity so unbiasedness is ensured asymptotically and we can use MSE for confidence intervals.} In this case we set $N=150$ and $T=200$. The idiosyncratic components are generated by Gaussian white noises cross-sectionally heteroscedastic and correlated. 

\begin{figure}[ht!]
\centering
\begin{tabular}{ccc}
\includegraphics[width = 0.3 \textwidth]{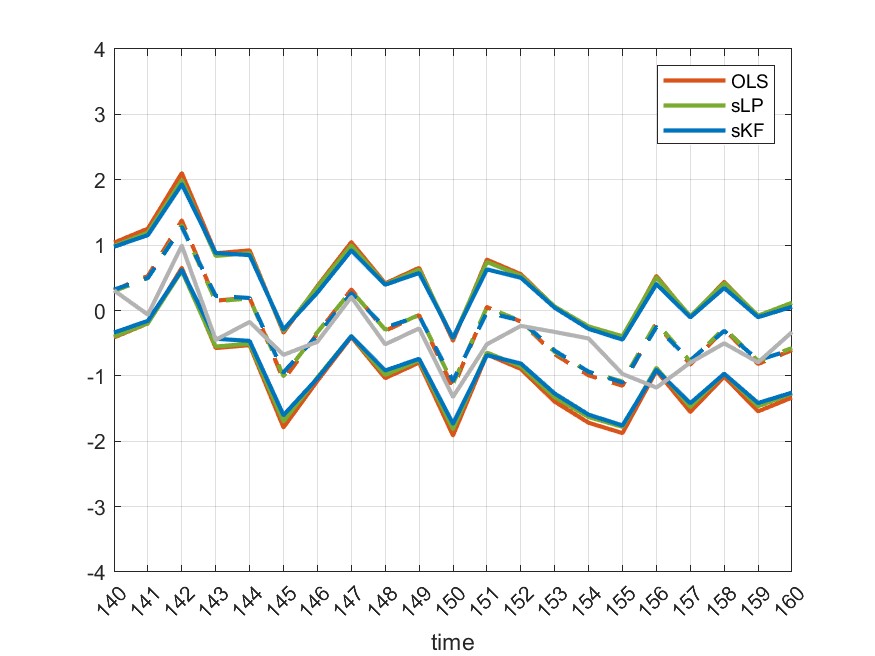}
\includegraphics[width = 0.3 \textwidth]{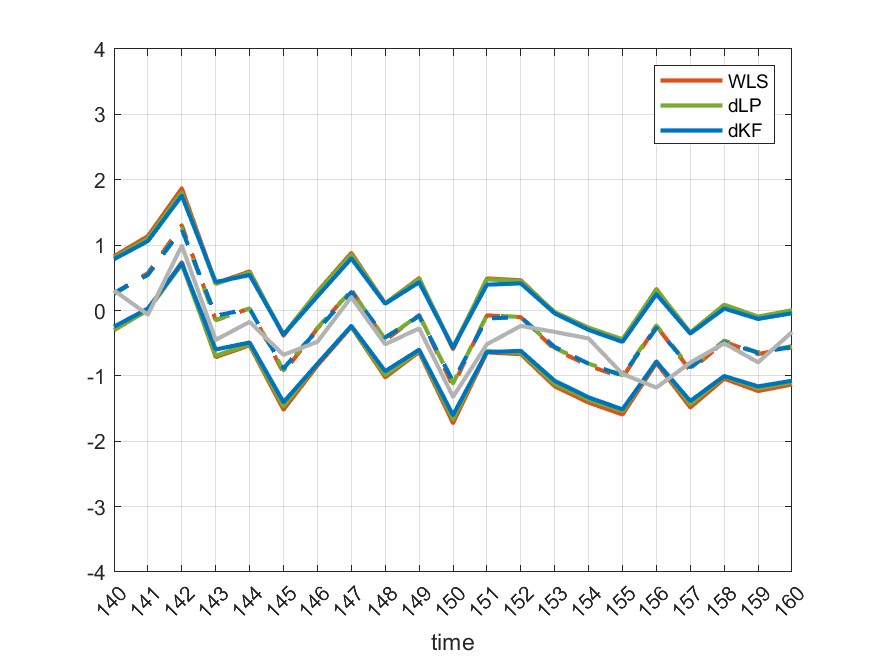}
\includegraphics[width = 0.3 \textwidth]{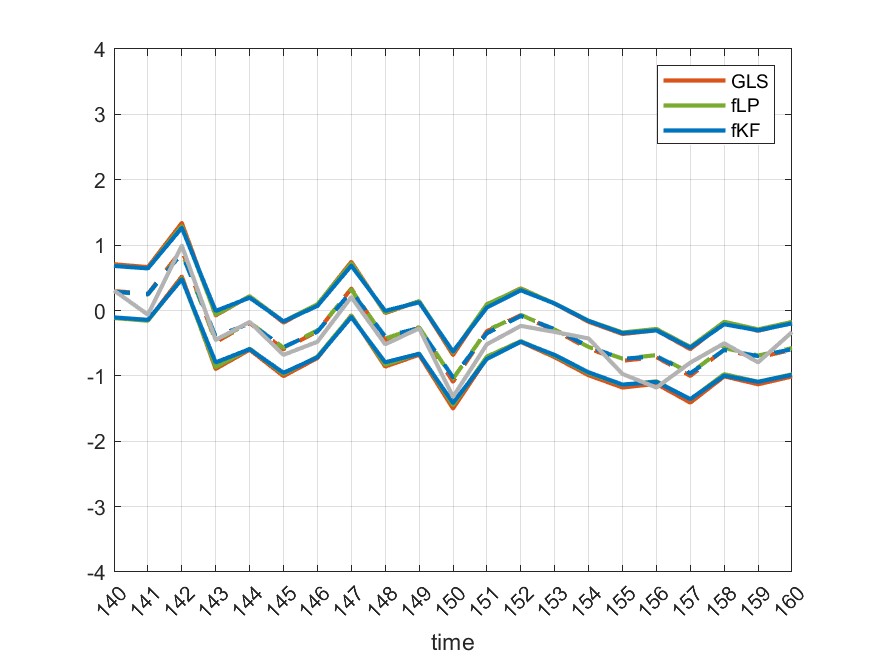}\\
\includegraphics[width = 0.3 \textwidth]{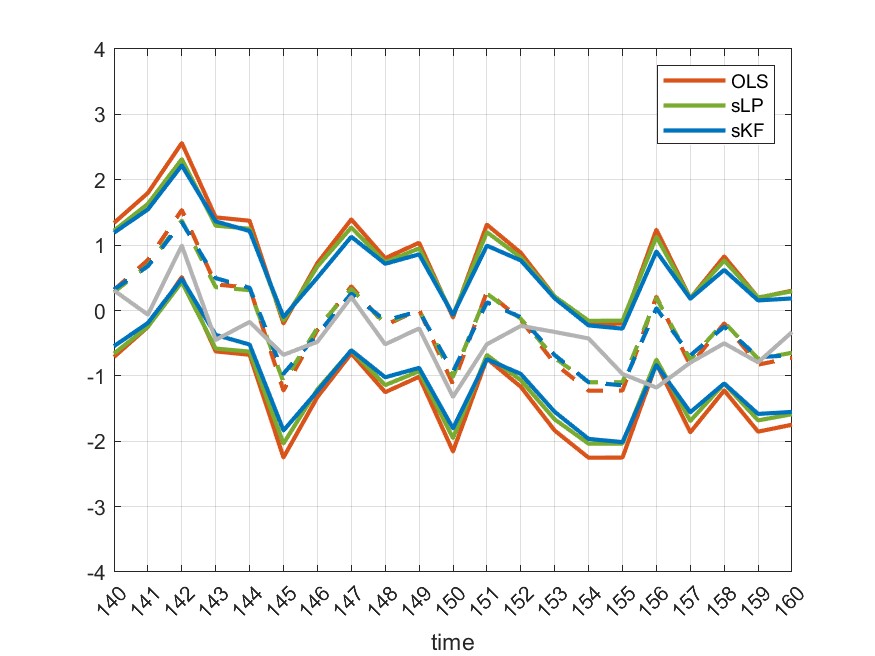}
\includegraphics[width = 0.3 \textwidth]{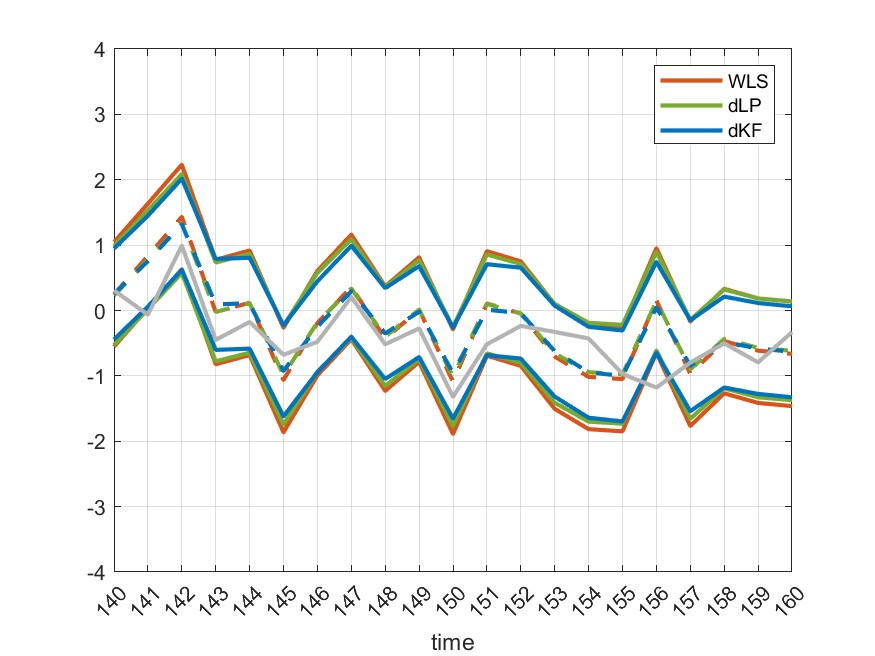}
\includegraphics[width = 0.3 \textwidth]{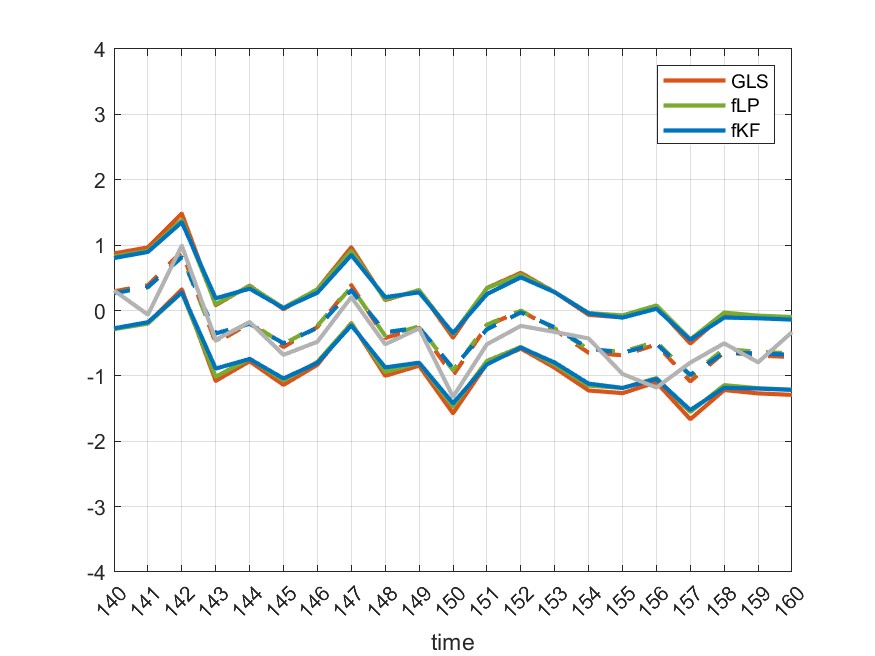}\\
\includegraphics[width = 0.3 \textwidth]{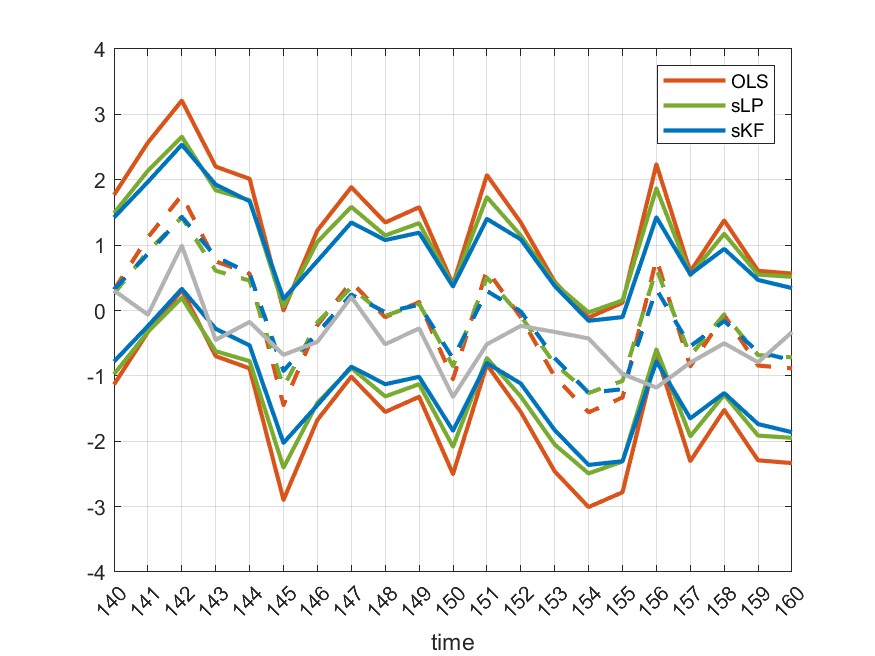}
\includegraphics[width = 0.3 \textwidth]{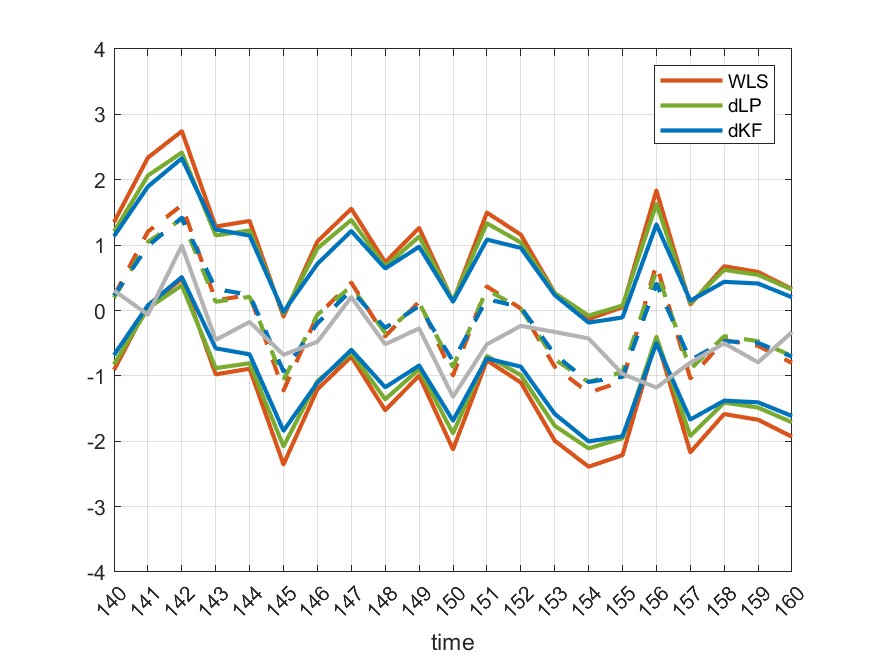}
\includegraphics[width = 0.3 \textwidth]{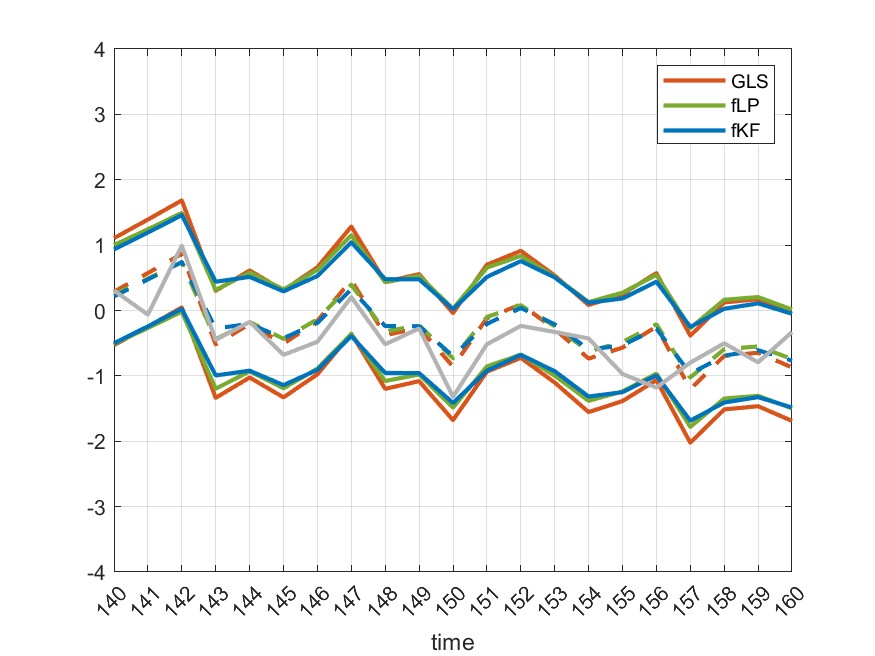} 
\end{tabular}

\caption{Simulated AR factor with $\phi=0.7$ (continuous grey line) from DFM with $N=150$ and $t=140,\ldots,160$, together with estimates (dashed lines) based on assuming scalar idiosyncratic covariance matrix (first column), diagonal idiosyncratic covariance matrix (second column), and full idiosyncratic covariance matrix (third column) when the true covariance matrix is full. The noise-to-signal ratio is: $\sigma^{2*}=0.5$ (first row), $\sigma^{2*}=1$ (second row), and $\sigma^{2*}=2$ (third row). Pairwise 95\% confidence bounds (continuous lines).}
\label{fig:factor3}
\end{figure}

Figure \ref{fig:factor3} plots the true simulated factor when $\phi=0.7$ together with its estimates and corresponding 95\% asymptotic confidence bands when the idiosyncratic covariance matrix is full with different noise-to-signal ratios. The MSEs are computed assuming scalar, diagonal and full covariance matrices. This figure illustrates the lost of efficiency when using procedures that wrongly assume scalar or diagonal idiosyncratic covariance matrix. We can observe that the reduction of MSE obtained when using extraction procedures that assume full idiosyncratic covariance can be large when $\mathbf {\Sigma}_{\varepsilon}$ is truly full. Furthermore, it also illustrates that the MSEs are obviously larger the larger the variance of the idiosyncratic noises. The differences between the estimated factors and their corresponding pairwise confidence bounds obtained by using LP or KF are indistinguishable, and LS is only slightly less efficient when $\sigma^{2*}=2$.


\begin{figure}
\centering
\begin{tabular}{ccc}
\includegraphics[width = 0.3 \textwidth]{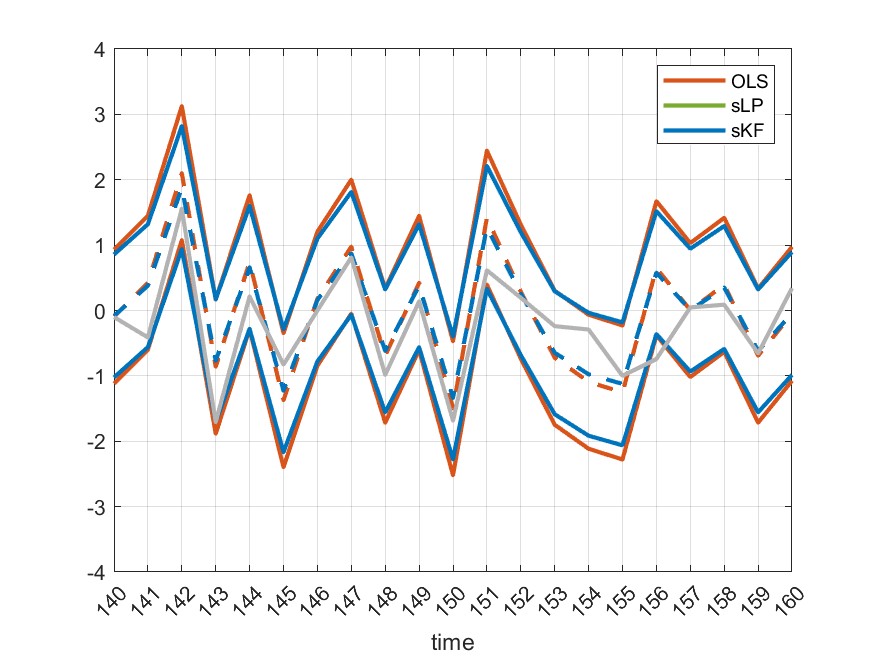}
\includegraphics[width = 0.3 \textwidth]{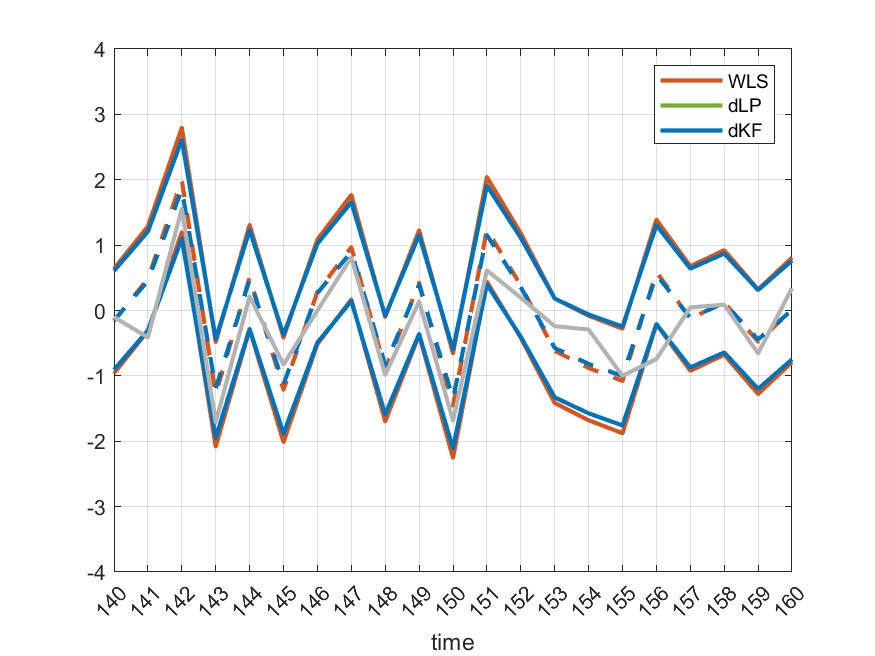}
\includegraphics[width = 0.3 \textwidth]{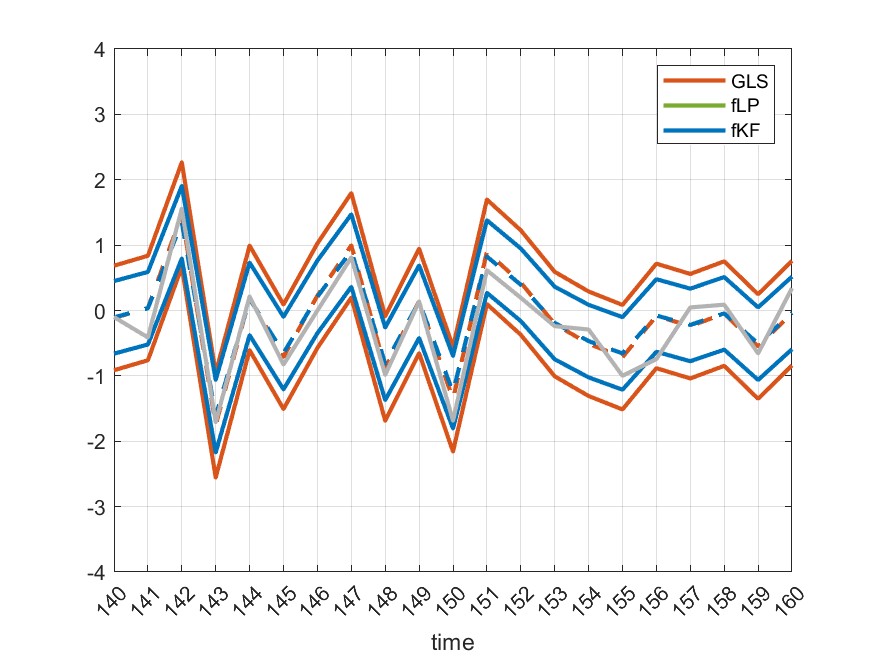}\\
\includegraphics[width = 0.3 \textwidth]{Scalar_sig_ast1_phi07}
\includegraphics[width = 0.3 \textwidth]{Diagonal_sig_ast1_phi07}
\includegraphics[width = 0.3 \textwidth]{Full_sig_ast1_phi07}\\
\includegraphics[width = 0.3 \textwidth]{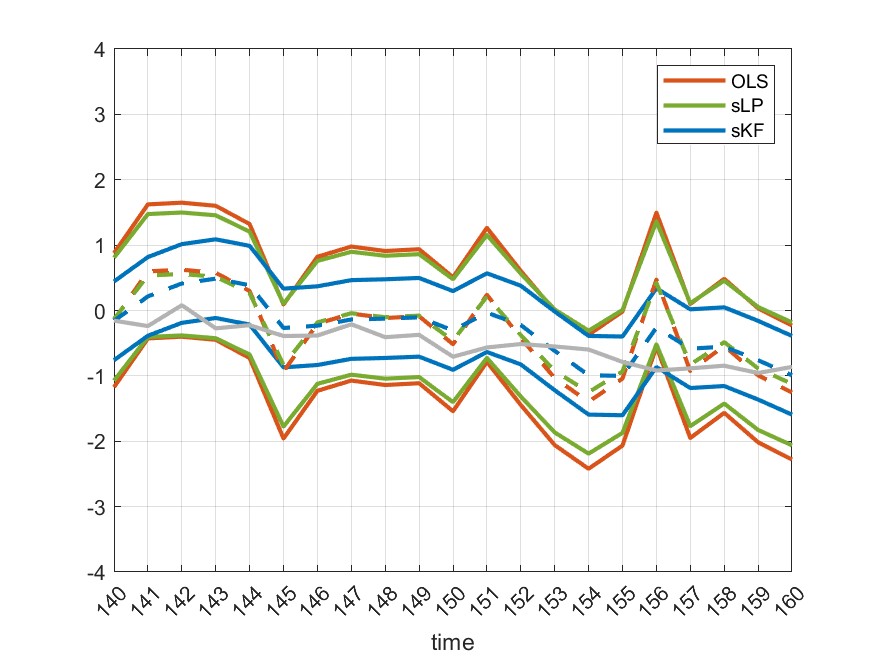}
\includegraphics[width = 0.3 \textwidth]{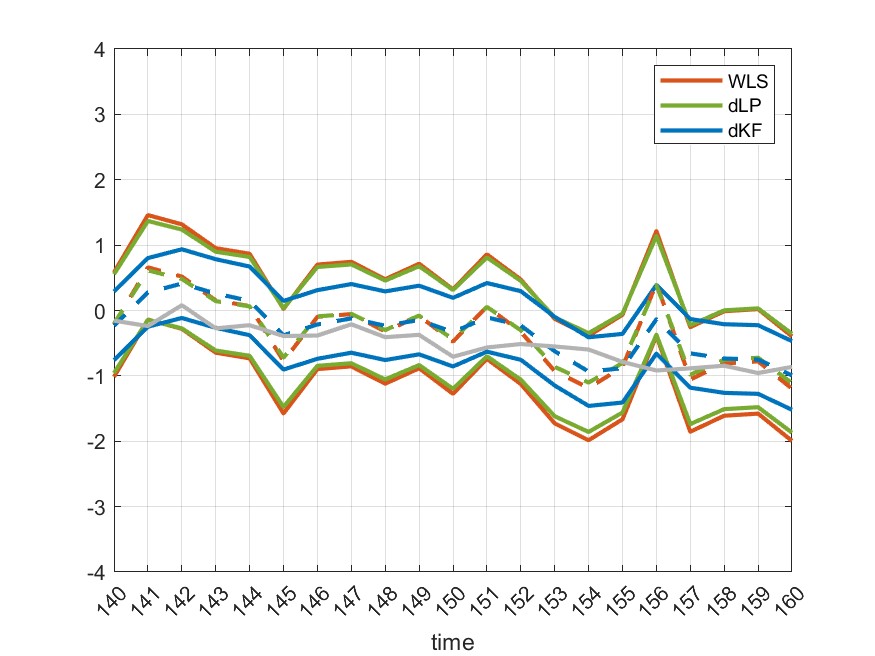}
\includegraphics[width = 0.3 \textwidth]{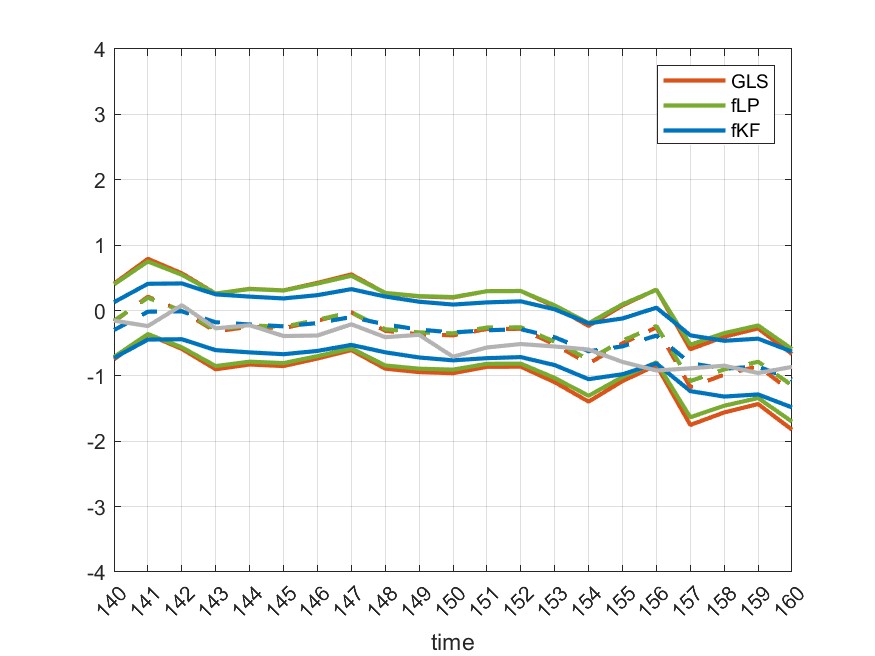}
\end{tabular}

\caption{Simulated AR factor (continuous grey line) with $\phi=0$ (first row), $\phi=0.7$ (second row) and $\phi=0.97$ (third row), from DFM with $N=150$ and $t=140,\ldots,160$, together with estimates (dashed lines) based on assuming scalar idiosyncratic covariance matrix (first column), diagonal idiosyncratic covariance matrix (second column), and full idiosyncratic covariance matrix (third column) when the true covariance matrix is full. The noise-to-signal ratio is $\sigma^{2*}=1$. Pairwise 95\% confidence bounds (continuous lines).}
\label{fig:illustra}
\end{figure}

The role of the persistence of the temporal dependence of the factors on the above results is illustrated in Figure \ref{fig:illustra}, in which, together with the true factors simulated with $\phi=0, 0.7$ and 0.97, we plot their estimates and $95\%$ asymptotic confidence bands. In this case, the noise-to-signal ratio is $\sigma^{2*}=1$. We can see that the differences between LP  and KF estimates of the factors can be huge when the factor is close to non-stationarity, i.e., when dynamics matters more.

\color{black}\section{Conclusions}
\label{sec:final}

In DFMs, there are two conceptually different approaches to estimating latent factors. The first considers that factors are fixed, treating them as unknown constants (or parameters) to be estimated. This is the target of PC factor extraction by Bai (2003), where the goal is to estimate the actual realizations of the factors. The second approach treats the factors as serially independent or autocorrelated random variables and estimates their linear projections given the contemporaneous or lagged observed data, respectively, as in the KF framework. Although it may seem counter-intuitive, the MSEs of the estimated factors are lower under the stochastic factor assumption. This is because the MSE in the random case reflects uncertainty about the linear projection of the factor, not about a specific realization, which is inherently harder to estimate precisely in the presence of noise. The difference stems not from the amount of information available, but from what is being estimated (a linear projection versus a fixed unknown constant).

We also show that the MSEs of the KF factors extracted using only contemporaneous observations or also lagged observations are very similar unless the serial dependence of the factors is very persistent, implying close to non-stationary factors. Finally, we study and compare the MSEs of PC and KF factors when the idiosyncratic components are wrongly assumed to be cross-sectionally homoscedastic and/or uncorrelated. 

The relevance of the results for the construction of confidence bounds for the factors are illustrated with simulated data. 

\section*{Acknowledgements}

The third author acknowledges partial financial support from the Spanish National Research Agency (Ministry of Science and Technology) Project PID2022-139614NB-C22. Part of this research was carried out while the third author was visiting the Department of Economics of the University of Bologna. She thanks the department for its hospitality and acknowledges financial support from the Salvador Madariaga grant PRX22-00195 for this visit. Any remaining errors are obviously our responsibility.

\newpage

\appendix

\section*{Appendix}

\subsection*{Appendix A: Convergence of MSE of Linear Projections}
\label{sec:Appendix_C}
\setcounter{equation}{0}
\renewcommand{\theequation}{A.\arabic{equation}}
\label{app:C}

Here we prove that the MSEs of dLP and sLP converge to zero as $N\to\infty$ and that the dLP (sLP) is asymptotically equivalent to the WLS (OLS) estimator.

First, consider the MSE of dLP in \eqref{eq:MSE_DLP_2}. Let define $\mathbf{M}=\left(\mathbf{\Lambda}\mathbf{\Lambda}^{\prime}+ \mathbf{\Sigma_{\varepsilon}}^{\ast}\right)^{-1}\mathbf{\Lambda}$, so that
\begin{equation}
\text{MSE}(\mathbf{f}_{t}^{dLP})
=
\mathbf{I}_{r}
+ \mathbf{M}^{\prime}
(\mathbf{\Lambda}\mathbf{\Lambda}^{\prime}
+ \mathbf{\Sigma_{\varepsilon}})
\mathbf{M}
- 2\mathbf{\Lambda}^{\prime}\mathbf{M}.
\label{eq:MSE_DLP_A}
\end{equation}
Using the Woodbury matrix identity
\[
(\mathbf{A} + \mathbf{UCV})^{-1}
= \mathbf{A}^{-1}
- \mathbf{A}^{-1}\mathbf{U}(\mathbf{C}^{-1}+\mathbf{VA}^{-1}\mathbf{U})^{-1}\mathbf{VA}^{-1},
\]
with
$\mathbf{A} = \mathbf{\Sigma}_{\varepsilon}^{\ast}$,
$\mathbf{U} = \mathbf{\Lambda}$,
$\mathbf{C} = \mathbf{I}_{r}$,
and $\mathbf{V} = \mathbf{\Lambda}^{\prime}$,
we obtain
\begin{equation}
(\mathbf{\Lambda}\mathbf{\Lambda}^{\prime}
+ \mathbf{\Sigma}_{\varepsilon}^{\ast})^{-1}
=
\mathbf{\Sigma}_{\varepsilon}^{\ast-1}
- \mathbf{\Sigma}_{\varepsilon}^{\ast-1}
\mathbf{\Lambda}
(\mathbf{I}_{r}
+ \mathbf{\Lambda}^{\prime}
\mathbf{\Sigma}_{\varepsilon}^{\ast-1}
\mathbf{\Lambda})^{-1}
\mathbf{\Lambda}^{\prime}
\mathbf{\Sigma}_{\varepsilon}^{\ast-1}.
\end{equation}
Therefore,
\begin{align}
\mathbf{M}
&=
\mathbf{\Sigma}_{\varepsilon}^{\ast-1}\mathbf{\Lambda}
-
\mathbf{\Sigma}_{\varepsilon}^{\ast-1}\mathbf{\Lambda}
(\mathbf{I}_{r}
+ \mathbf{\Lambda}^{\prime}
\mathbf{\Sigma}_{\varepsilon}^{\ast-1}
\mathbf{\Lambda})^{-1}
\mathbf{\Lambda}^{\prime}
\mathbf{\Sigma}_{\varepsilon}^{\ast-1}\mathbf{\Lambda}
\nonumber\\
&=
\mathbf{\Sigma}_{\varepsilon}^{\ast-1}\mathbf{\Lambda}
\!\left[
\mathbf{I}_{r}
- (\mathbf{I}_{r}
+ \mathbf{\Lambda}^{\prime}
\mathbf{\Sigma}_{\varepsilon}^{\ast-1}
\mathbf{\Lambda})^{-1}
\mathbf{\Lambda}^{\prime}
\mathbf{\Sigma}_{\varepsilon}^{\ast-1}\mathbf{\Lambda}
\right].
\end{align}
Using now
\[
\mathbf{A}(\mathbf{A}+\mathbf{I})^{-1} = (\mathbf{A}+\mathbf{I})^{-1}\mathbf{A} =\mathbf{ I} - (\mathbf{A}+\mathbf{I})^{-1},
\]
with
$\mathbf{A} = \mathbf{\Lambda}^{\prime}\mathbf{\Sigma}_{\varepsilon}^{\ast-1}\mathbf{\Lambda}$, we obtain
\begin{align}
\mathbf{M}
&=
\mathbf{\Sigma}_{\varepsilon}^{\ast-1}\mathbf{\Lambda}
(\mathbf{I}_{r}
+ \mathbf{\Lambda}^{\prime}
\mathbf{\Sigma}_{\varepsilon}^{\ast-1}
\mathbf{\Lambda})^{-1}.
\label{eq:M_NVYA_N}
\end{align}
Substituting equation \eqref{eq:M_NVYA_N} into the second term of \eqref{eq:MSE_DLP_A}, we get
\begin{align}
\mathbf{M}^{\prime}
(\mathbf{\Lambda}\mathbf{\Lambda}^{\prime}
+ \mathbf{\Sigma_{\varepsilon}})
\mathbf{M}
&=
(\mathbf{I}_{r}
+ \mathbf{\Lambda}^{\prime}
\mathbf{\Sigma}_{\varepsilon}^{\ast-1}
\mathbf{\Lambda})^{-1}
\mathbf{\Lambda}^{\prime}
\mathbf{\Sigma}_{\varepsilon}^{\ast-1}
(\mathbf{\Lambda}\mathbf{\Lambda}^{\prime}
+ \mathbf{\Sigma_{\varepsilon}})
\mathbf{\Sigma}_{\varepsilon}^{\ast-1}
\mathbf{\Lambda}
(\mathbf{I}_{r}
+ \mathbf{\Lambda}^{\prime}
\mathbf{\Sigma}_{\varepsilon}^{\ast-1}
\mathbf{\Lambda})^{-1}
\nonumber\\
&=
(\mathbf{I}_{r}
+ \mathbf{\Lambda}^{\prime}
\mathbf{\Sigma}_{\varepsilon}^{\ast-1}
\mathbf{\Lambda})^{-1}
\mathbf{\Lambda}^{\prime}
\mathbf{\Sigma}_{\varepsilon}^{\ast-1}\mathbf{\Lambda}
\mathbf{\Lambda}^{\prime}
\mathbf{\Sigma}_{\varepsilon}^{\ast-1}\mathbf{\Lambda}
(\mathbf{I}_{r}
+ \mathbf{\Lambda}^{\prime}
\mathbf{\Sigma}_{\varepsilon}^{\ast-1}
\mathbf{\Lambda})^{-1}
\nonumber\\
&\quad +
(\mathbf{I}_{r}
+ \mathbf{\Lambda}^{\prime}
\mathbf{\Sigma}_{\varepsilon}^{\ast-1}
\mathbf{\Lambda})^{-1}
\mathbf{\Lambda}^{\prime}
\mathbf{\Sigma}_{\varepsilon}^{\ast-1}
\mathbf{\Sigma}_{\varepsilon}
\mathbf{\Sigma}_{\varepsilon}^{\ast-1}
\mathbf{\Lambda}
(\mathbf{I}_{r}
+ \mathbf{\Lambda}^{\prime}
\mathbf{\Sigma}_{\varepsilon}^{\ast-1}
\mathbf{\Lambda})^{-1}.
 \label{eq:M_NVYA_M}
\end{align}
The first term in the last expression can be writeen as
\begin{equation}
\label{eq:first_term}
\left(
\frac{\mathbf{I}_{r}}{N}
+ \frac{\mathbf{\Lambda}^{\prime}\mathbf{\Sigma}_{\varepsilon}^{\ast-1}\mathbf{\Lambda}}{N}
\right)^{-1}
\left(
\frac{\mathbf{\Lambda}^{\prime}\mathbf{\Sigma}_{\varepsilon}^{\ast-1}\mathbf{\Lambda}}{N}
\right)
\left(
\frac{\mathbf{\Lambda}^{\prime}\mathbf{\Sigma}_{\varepsilon}^{\ast-1}\mathbf{\Lambda}}{N}
\right)
\left(
\frac{\mathbf{I}_{r}}{N}
+ \frac{\mathbf{\Lambda}^{\prime}\mathbf{\Sigma}_{\varepsilon}^{\ast-1}\mathbf{\Lambda}}{N}
\right)^{-1}.
\end{equation}
We observe that, as $N \to \infty$, the term $\mathbf{I}_{r}/N$ becomes negligible 
compared to $\mathbf{\Lambda}^{\prime}\mathbf{\Sigma}_{\varepsilon}^{\ast-1}\mathbf{\Lambda}/N$. 
Thus,
\begin{equation}
\label{eq:approx_inv}
\left(
\frac{\mathbf{I}_{r}}{N}
+ \frac{\mathbf{\Lambda}^{\prime}\mathbf{\Sigma}_{\varepsilon}^{\ast-1}\mathbf{\Lambda}}{N}
\right)^{-1}
=
\left(
\frac{\mathbf{\Lambda}^{\prime}\mathbf{\Sigma}_{\varepsilon}^{\ast-1}\mathbf{\Lambda}}{N}
\right)^{-1}
+ O\!\left(\frac{1}{N}\right).
\end{equation}
On the other hand, the second term in  \eqref{eq:M_NVYA_M} can be written as
\begin{equation}
\label{eq:second_term}
\frac{1}{N}
\left(
\frac{\mathbf{I}_{r}}{N}
+ \frac{\mathbf{\Lambda}^{\prime}\mathbf{\Sigma}_{\varepsilon}^{\ast-1}\mathbf{\Lambda}}{N}
\right)^{-1}
\left(
\frac{\mathbf{\Lambda}^{\prime}\mathbf{\Sigma}_{\varepsilon}^{\ast-1}
\mathbf{\Sigma}_{\varepsilon}
\mathbf{\Sigma}_{\varepsilon}^{\ast-1}\mathbf{\Lambda}}{N}
\right)
\left(
\frac{\mathbf{I}_{r}}{N}
+ \frac{\mathbf{\Lambda}^{\prime}\mathbf{\Sigma}_{\varepsilon}^{\ast-1}\mathbf{\Lambda}}{N}
\right)^{-1}=O\left(\frac 1N\right),
\end{equation}
As a result,
\begin{equation}
\label{eq:first_term_limit}
\mathbf M^{\prime}
(\mathbf{\Lambda}\mathbf{\Lambda}^{\prime}
+ \mathbf{\Sigma}_{\varepsilon})
\mathbf M
=
\mathbf{I}_{r} + O\left(\frac 1N\right).
\end{equation}
Finally, note that the third term in  \eqref{eq:MSE_DLP_A}  is
\begin{align}
\label{eq:LambdaM}
\mathbf{\Lambda}^{\prime}\mathbf{M}
&=
\mathbf{\Lambda}^{\prime}\mathbf{\Sigma}_{\varepsilon}^{\ast-1}\mathbf{\Lambda}
\Big[
\mathbf{I}_{r}
- (\mathbf{I}_{r}
+ \mathbf{\Lambda}^{\prime}\mathbf{\Sigma}_{\varepsilon}^{\ast-1}\mathbf{\Lambda})^{-1}
\mathbf{\Lambda}^{\prime}\mathbf{\Sigma}_{\varepsilon}^{\ast-1}\mathbf{\Lambda}
\Big]
\nonumber\\
&=
\mathbf{\Lambda}^{\prime}\mathbf{\Sigma}_{\varepsilon}^{\ast-1}\mathbf{\Lambda}
(\mathbf{I}_{r}
+ \mathbf{\Lambda}^{\prime}\mathbf{\Sigma}_{\varepsilon}^{\ast-1}\mathbf{\Lambda})^{-1}
\nonumber\\
&=
\frac{\mathbf{\Lambda}^{\prime}\mathbf{\Sigma}_{\varepsilon}^{\ast-1}\mathbf{\Lambda}}{N}
\left(
\frac{\mathbf{I}_{r}}{N}
+ \frac{\mathbf{\Lambda}^{\prime}\mathbf{\Sigma}_{\varepsilon}^{\ast-1}\mathbf{\Lambda}}{N}
\right)^{-1}
= \mathbf{I}_{r} + O\left(\frac 1N\right).
\end{align}
Combining results from 
\eqref{eq:first_term_limit} and \eqref{eq:LambdaM}, 
we conclude that
\begin{equation}
\label{eq:MSE_limit_result}
\text{MSE}(\mathbf{f}_{t}^{dLP})
=
\mathbf{I}_{r}
+ \mathbf{I}_{r}
- 2\mathbf{I}_{r}
+O\left(\frac 1N\right)=O\left(\frac 1N\right),
\end{equation}
thus, the MSE of the dLP estimator converges to zero as $N\to\infty$.

An analogous argument applies to the case in which the idiosyncratic covariance matrix is 
misspecified as being spherical.
Indeed, by replacing $\mathbf{\Sigma}_{\varepsilon}^{\ast}$ by $\sigma_{\varepsilon}^{2}\mathbf{I}_N$ in the expressions above
and following the same steps, one obtains again that
\begin{equation}\label{eq:MSE_limit_result_sLP}
\text{MSE}(\mathbf{f}_{t}^{sLP}) =O\left(\frac 1N\right).
\end{equation}


Second, using a first-order asymptotic expansion in powers of 
$\mathbf{\Omega}^{-1}$, with 
$\mathbf{\Omega} = \mathbf{\Lambda}'\mathbf{\Sigma}_{\varepsilon}^{\ast-1}\mathbf{\Lambda}$,
we can rewrite
\begin{equation}
\label{eq:Omega_expand}
(\mathbf{I}_r + \mathbf{\Omega})^{-1}
= \mathbf{\Omega}^{-1} (\mathbf{I}_r + \mathbf{\Omega}^{-1})^{-1}.
\end{equation}
Since $\|\mathbf{\Omega}^{-1}\| = O(N^{-1}) \to 0$ as $N \to \infty$,
we can expand $(\mathbf{I}_r + \mathbf{\Omega}^{-1})^{-1}$ in powers of $\mathbf{\Omega}^{-1}$ as
\begin{eqnarray}
(\mathbf{I}_r + \mathbf{\Omega}^{-1})^{-1}
&=& \mathbf{I}_r - \mathbf{\Omega}^{-1} + \mathbf{\Omega}^{-2} + O(\mathbf{\Omega}^{-3}). 
\label{eq:Omega_series}
\end{eqnarray}
Thus,
\begin{equation}
\label{eq:Omega_asymptotic}
(\mathbf{I}_r + \mathbf{\Omega})^{-1}
= \mathbf{\Omega}^{-1} - \mathbf{\Omega}^{-2} + O(\mathbf{\Omega}^{-3}),
\end{equation}
which holds because $\mathbf{\Omega}$ is symmetric positive definite and of order $O(N)$,
so $\mathbf{\Omega}^{-1}$ becomes asymptotically negligible as $N$ grows. Replacing \eqref{eq:Omega_asymptotic} in the expression for the dLP estimator, it follows that
\begin{equation}\label{eq:MSE_dLP_TAYLOR}
\begin{aligned}
\mathbf{f}_t^{dLP}
&=
\Big[
\mathbf{\Omega}^{-1}
- \mathbf{\Omega}^{-2}
+ O(N^{-3})
\Big]
\mathbf{\Lambda}'\mathbf{\Sigma}_{\varepsilon}^{\ast-1}\mathbf{Y}_t\\[3pt]
&=
\mathbf{\Omega}^{-1}\mathbf{\Lambda}'\mathbf{\Sigma}_{\varepsilon}^{\ast-1}\mathbf{Y}_t
- \mathbf{\Omega}^{-2}\mathbf{\Lambda}'\mathbf{\Sigma}_{\varepsilon}^{\ast-1}\mathbf{Y}_t
+ O_p(N^{-2}),
\end{aligned}
\end{equation}
since $\mathbf{\Lambda}'\mathbf{\Sigma}_{\varepsilon}^{\ast-1}\mathbf{Y}_t=O(N)$. The first term coincides with the WLS estimator, 
while the remaining terms are of higher order in $N^{-1}$ 
and vanish asymptotically as $N \to \infty$. 
In particular,
\begin{equation}\label{eq:MSE_dLP_vs_WLS}
\begin{split}
&\mathbf{f}_t^{dLP} 
= \mathbf{f}_t^{WLS} +O_p\left(\frac 1N\right),\\
&N\text{MSE}(\mathbf{f}_t^{dLP}) - N\text{MSE}(\mathbf{f}_t^{WLS})= O\left(\frac 1N\right).
\end{split}
\end{equation}
Since the WLS estimator is $\sqrt N$--consistent and asymptotically normal,
Slutsky's theorem implies that the dLP estimator is also $\sqrt N$--consistent and asymptotically normal
with the same distribution as the WLS estimator.\\

A similar line of reasoning can be applied to the sLP estimator, when replacing $\mathbf{\Sigma}_{\varepsilon}^{\ast}$ with $\sigma_{\varepsilon}^{2}\mathbf{I}_N$ in the expressions above.  Thus,
\begin{equation}\label{eq:MSE_sLP_vs_OLS}
\begin{split}
&\mathbf{f}_t^{sLP} 
= \mathbf{f}_t^{OLS} +O_p\left(\frac 1N\right),\\
&N\text{MSE}(\mathbf{f}_t^{sLP}) - N\text{MSE}(\mathbf{f}_t^{OLS})= O\left(\frac 1N\right).
\end{split}
\end{equation}
Since the sLP and OLS estimators are asymptotically equivalent as $N \to \infty$,  
it follows that the sLP estimator is also $\sqrt N$--consistent and converges  
to the same asymptotic normal distribution as the OLS estimator.

\end{document}